\pdfoutput=1
\RequirePackage{ifpdf}
\ifpdf 
\documentclass[pdftex]{sigma}
\else
\documentclass{sigma}
\fi

\DeclareMathOperator{\diag}{diag}
\numberwithin{equation}{section}

\begin{document}

\newcommand{\oo}[0]{\otimes}
\newcommand\adsw{AdS$_\omega$\;}
\def\pois#1#2{\left\{#1,#2 \right\}}

\allowdisplaybreaks

\renewcommand{\thefootnote}{$\star$}

\renewcommand{\PaperNumber}{052}

\FirstPageHeading

\ShortArticleName{Twisted (2+1) $\kappa$-AdS Algebra, Drinfel'd Doubles and Non-Commutative Spacetimes}

\ArticleName{Twisted (2+1) $\boldsymbol{\kappa}$-AdS Algebra, Drinfel'd Doubles
\\
and Non-Commutative Spacetimes\footnote{This paper is a~contribution to the Special Issue on Deformations of Space-Time
and its Symmetries.
The full collection is available at \href{http://www.emis.de/journals/SIGMA/space-time.html}
{http://www.emis.de/journals/SIGMA/space-time.html}}}

\Author{\'Angel BALLESTEROS~$^\dag$, Francisco J.~HERRANZ~$^\dag$, Catherine MEUSBURGER~$^\ddag$ and\\
Pedro NARANJO~$^\dag$}
\AuthorNameForHeading{\'A.~Ballesteros, F.J.~Herranz, C.~Meusburger and P.~Naranjo}

\Address{$^\dag$~Departamento de F{\'{\i}}sica, Universidad de Burgos, E-09001 Burgos, Spain}

\EmailD{\href{mailto:angelb@ubu.es}{angelb@ubu.es}, \href{mailto:fjherranz@ubu.es}{fjherranz@ubu.es},
\href{mailto:pnaranjo@ubu.es}{pnaranjo@ubu.es}}

\Address{$^\ddag$~Department Mathematik, Friedrich-Alexander Universit\"at Erlangen-N\"urnberg,\\
\hphantom{$^\ddag$}~Cauerstr.~11, D-91058 Erlangen, Germany}

\EmailD{\href{mailto:catherine.meusburger@math.uni-erlangen.de}{catherine.meusburger@math.uni-erlangen.de}}

\ArticleDates{Received March 09, 2014, in f\/inal form May 13, 2014; Published online May 18, 2014}

\Abstract{We construct the full quantum algebra, the corresponding Poisson--Lie structure and the associated quantum
spacetime for a~family of quantum deformations of the isometry algebras of the (2+1)-dimensional anti-de Sitter (AdS),
de Sitter (dS) and Minkowski spaces.
These deformations correspond to a~Drinfel'd double structure on the isometry algebras that are motivated by their role
in (2+1)-gravity.
The construction includes the cosmological constant $\Lambda$ as a~deformation parameter, which allows one to treat
these cases in a~common framework and to obtain a~twisted version of both space- and time-like $\kappa$-AdS and dS
quantum algebras; their f\/lat limit $\Lambda\to 0$ leads to a~twisted quantum Poincar\'e algebra.
The resulting non-commutative spacetime is a~nonlinear $\Lambda$-deformation of the $\kappa$-Minkowski one plus an
additional contribution generated by the twist.
For the AdS case, we relate this quantum deformation to two copies of the standard (Drinfel'd--Jimbo) quantum
deformation of the Lorentz group in three dimensions, which allows one to determine the impact of the twist.}


\Classification{16T20; 81R50; 81R60}

\renewcommand{\thefootnote}{\arabic{footnote}} \setcounter{footnote}{0}

\section{Introduction}

Spacetime geometry is widely expected to exhibit discrete features at the Planck scale~\mbox{\cite{Garay, Majida}}, which gives
rise to an interplay between algebra and geometry.
A schematic approach to these phenomena is provided by non-commutative models, in which spacetime coordinates are
replaced by (non-commuting) operators whose uncertainty relations encode the discrete nature of spacetime geometry.

Non-commutative models based on quantum groups~\cite{CP, majid} are well-established in this context and serve as
a~framework which enables one to construct non-commutative spacetimes together with an action of the corresponding
generalisations of the classical spacetime kinematical groups~\cite{BLL}.
Most of these models are based on~$q$-deformed function algebras and the associated universal enveloping algebras, where
the deformation parameter~$q$ is related to the Planck scale.

Since the introduction of the so-called $\kappa$-Poincar\'e algebras and their associated $\kappa$-Minkowski
spacetimes~\cite{bicross,RoxPLB,CK4,CK3,nullplaneA,nullplaneB,Borowiec,Borosigma, Dasz1,Dasz2, azc,AP,Lukierskicd,Lukierskicc,Lukierskib,LukR,Lukierskia,LukNR,Majid:1994cy,Maslanka,kZakr},
dif\/ferent quantum deformations of kinematical groups of spacetimes have been proposed, and a~large number of possible
deformations has been outlined.
The resulting $\kappa$-Poincar\'e models played an important role in the development of the so-called ``doubly special
relativity'' theories~\cite{Amelino-Camelia:2000mrr,AmelinoLRR,amel,Brunob,Brunoa,KowalskiFS,
Kowalski-Glikman:2002we,Lukierski:2002df, MagueijoSmolin}, where the deformation
parameter $\kappa$ is interpreted as a~second fundamental constant (related to the Plank length) in addition to the
speed of light.

However, most of the work in this context focused on deformations of Minkowski space and the Poincar\'e group, while
quantum anti-de Sitter (AdS) space and de Sitter (dS) space and the associated isometry groups have received less
attention (see~\cite{Rox,CK3,Marciano} and references therein).
This is regrettable since it does not allow one to investigate ef\/fects due to a~non-trivial cosmological constant or,
more generally, non-trivial curvature in these models.
Indeed, the interplay between curvature and Planck scale ef\/fects should be taken into account when the possible
implications in astrophysics and cosmology of the novel properties of spacetime at the Planck scale are
considered~\cite{AmelinoLRR, Marciano}.

Since quantum deformations are highly non-unique, another essential open problem is the question regarding which of the
models under consideration is suitable for the description of ef\/fects at the Planck scale.
This question is relevant since even quantum deformations related by a~twist give rise to dif\/ferent composition laws for
multi-particle systems.
However, in (3+1) dimensions, arguments about the physical interpretation of models generated by dif\/ferent quantum
deformations are largely heuristic and phenomenological.

In contrast, the (2+1)-dimensional scenario of\/fers some insight into this question, as quantum group structures arise
naturally in the quantisation of (2+1)-gravity.
They can be viewed as the quantum counterparts of certain Poisson--Lie symmetries which describe the Poisson structure
on the phase space of the theory.
This restricts the possible quantum deformations that are compatible with classical (2+1)-gravity~\cite{cm2}.
It was shown in~\cite{BHMcqg} that a~criterium that ensures compatibility with (2+1)-gravity is the requirement that the
deformation arise from an underlying Drinfel'd double structure on the isometry group of spacetime.
Further evidence for the relevance of Drinfel'd doubles is their role in state sum or spin foam models of spacetimes,
which are related to the Turaev--Viro invariant~\cite{BK, TVir}.

A classif\/ication of the Drinfel'd double structures on the isometry groups of the classical homogeneous spacetimes can
therefore shed light on the question of which quantum deformations are relevant to the quantisation of (2+1)-gravity.
A further desirable property is that the possible deformations form a~family that admit the cosmological constant as
a~deformation parameter and hence permits one to include curvature ef\/fects in these models.

Based on these considerations, it was shown in~\cite{BHMcqg} that the isometry algebra of the (2+1)-dimensional AdS
space admits {\em two} classical~$r$-matrices arising from a~Drinfel'd double structure that are compatible with
(2+1)-gravity and its pairing~\cite{Witten1}, that determines its Poisson structure.
The quantum deformation associated with the f\/irst one was already presented in~\cite{BHM}, and the dependence of the
associated non-commutative spacetime on the cosmological constant was recently studied in~\cite{BHMletter}.
The second classical~$r$-matrix turns out to be a~twist of the quantum $\kappa$-AdS algebra~\cite{CK3} that, to our
knowledge, has not been previously considered in the literature, except for partial results.
In particular, its Poincar\'e counterpart was constructed in~\cite{Dasz1} and the corresponding pure $\kappa$-AdS
spacetime (without the twist) was considered in~\cite{Rox} and further studied in~\cite{Marciano}.

Moreover, it was shown in~\cite{BHMcqg} that both~$r$-matrices have counterparts for the dS and Poincar\'e algebras that
are compatible with (2+1)-gravity and give rise to quantum deformations of the isometry groups of these spacetimes.
This allows one to implement the cosmological constant as a~deformation parameter and to investigate the quantum ef\/fects
arising from curvature.

The aim of the present article is to construct the full quantum algebra for this twisted $\kappa$-deformation of the
AdS, dS and Poincar\'e algebras and to analyse the fundamental properties of their associated non-commutative
spacetimes.
We will construct these algebras simultaneously by considering the associated Lie algebras as a~family AdS$_\omega$, in
which the cosmological constant plays the role of a~real Lie algebra (graded contraction) parameter $\omega$.
The isometry algebras of AdS, dS and Minkowski spaces correspond, respectively, to positive, negative and zero values
of~$\omega$, which coincides with the {\em sectional curvature} of the underlying classical spacetime.

The structure of the paper is as follows.
In the next section we review the basic geometrical properties of (2+1)-dimensional AdS, dS and Minkowski spaces, their
isometry groups and the associated Lie algebras~\cite{sigma, conformes} and discuss how the latter can be grouped into
a~family AdS$_\omega$ for which the cosmological constant plays the role of a~deformation parameter $\omega$.
Section~\ref{Section3} exhibits the underlying Drinfel'd double structure that generates the twisted $\kappa$-AdS$_\omega$ via its
canonical classical~$r$-matrix.
We show that this Drinfel'd double structure leads to a~{\em space-like} $\kappa$-deformation~\cite{CK4, CK3} and that
the passage to the usual {\em time-like} one (i.e., the proper $\kappa$-deformation~\cite{LukNR}) involves a~change of
basis with complex coef\/f\/icients.
In Section~\ref{Section4} we compute the corresponding {\em first-order} structure of the full quantum deformation which is given by
its Lie bialgebra.
This Lie bialgebra depends on three {\em deformation} parameters $(\eta, z, \vartheta)$: the cosmological constant
$\Lambda =-\omega=-\eta^2$, the usual quantum (Plank scale) deformation parameter $\kappa=1/z$, and the additional
twist parameter $\vartheta$.
Furthermore, we construct and analyse the {\em first-order} non-commutative quantum spacetime for AdS$_\omega$, which
does not depend on the cosmological constant and hence is common to the three cases under consideration.

The construction of the full quantum twisted $\kappa$-AdS$_\omega$ algebra $U_{\kappa,\vartheta}(\mbox{AdS}_\omega)$ is
undertaken in Section~\ref{Section5}.
At this point, it is important to emphasise that the compatibility condition of the deformation with the underlying
Drinfel'd double structure is satisf\/ied only for very specif\/ic values of the deformation parameters $z=1/\kappa$ and
$\vartheta$, and this, in turn, depends on whether one chooses the space-like or time-like deformation from Section~\ref{Section3}.
We construct the (f\/lat) Poincar\'e limit $\omega\to0$ of the quantum algebra and show that this leads to one of the
known twisted $\kappa$-Poincar\'e algebras given in~\cite{Dasz1}.

In Section~\ref{Section6}, we construct the full Poisson--Lie group associated to this twisted deformation by making use of
a~simultaneous parametrisation of the (2+1) AdS, dS and Poincar\'e groups in terms of local coordinates.
The Poisson subalgebra generated by the local spacetime coordinates $\{x_0,x_1,x_2\}$ provides the classical counterpart
of the non-commutative twisted $\kappa$-AdS$_\omega$ spacetime associated to this deformation.
As expected, this Poisson AdS$_\omega$ spacetime is a~nonlinear algebra when the cosmological constant $\Lambda$ is
non-zero, and in the limit $\Lambda\to 0$ reduces to linear non-commutative Minkowski spacetime that is consistent with
the results in~\cite{Dasz1}.
Due to the nonlinear nature of this new AdS$_\omega$ spacetime, its quantisation is far from being trivial.
Never\-theless, in Section~\ref{Section7} it is shown that by using the Lie algebra isomorphism between $\mathfrak{so}(2,2)$ and
$\mathfrak{sl}(2,\mathbb{R})\oplus \mathfrak{sl}(2,\mathbb{R})$, the twisted $\kappa$-AdS Poisson--Lie group can be
reconstructed and fully quantised.
A f\/inal section with comments and a~discussion of open problems closes the paper.

\section[The AdS$_\omega$ Lie algebra]{The AdS$\boldsymbol{_\omega}$ Lie algebra}
\label{Section2}

In this section we describe, in a~unif\/ied setting, the realisation of the (2+1)-dimensional AdS, dS and Minkowski spaces
as symmetric homogeneous spaces, their isometry groups, their Lie algebras and the associated left- and right-invariant
vector f\/ields.
We start by considering their Lie algebras, which form a~family of three six-dimensional real Lie algebras which are
related by a~real contraction parameter $\omega$ and will be denoted by ${\rm AdS}_\omega \equiv
\mathfrak{so}_\omega(2,2)$ in the following.

In terms of a~basis $\{P_0, P_i, J, K_i\}$, $i=1,2$, consisting of the inf\/initesimal generators of, respectively, a~time
translation, spatial translations, a~spatial rotation and boosts, the Lie brackets of AdS$_\omega$ take the form
\begin{alignat}{5}
& [J,P_i]= \epsilon_{ij}P_j,
\qquad &&
[J,K_i]= \epsilon_{ij}K_j,
\qquad&&
[J,P_0]= 0,
\qquad&&
[P_i,K_j]=-\delta_{ij}P_0,&
\nonumber
\\
& [P_0,K_i]=-P_i,
\qquad&&
[K_1,K_2]=-J,
\qquad&&
[P_0,P_i]=\omega K_i,
\qquad&&
[P_1,P_2]=-\omega J,&
\label{ba}
\end{alignat}
where $i,j=1,2$ and $\epsilon_{ij}$ is a~skew-symmetric tensor normalised such that $\epsilon_{12}=1$.
For positive, zero and negative values of $\omega$, this Lie bracket def\/ines a~Lie algebra isomorphic to
$\mathfrak{so}(2,2)$, $\mathfrak{iso}(2,1)= \mathfrak{so}(2,1)\ltimes \mathbb R^3$ and $\mathfrak{so}(3,1)$,
respectively.
Note that when $\omega\ne 0$, this parameter can always be transformed to $\omega=\pm 1$ by a~rescaling of the basis.
Moreover, the case $\omega= 0$ can be understood as an In\"on\"u--Wigner contraction~\cite{IWa}: $\mathfrak{so}(2,2)\to
\mathfrak {iso}(2,1)\leftarrow \mathfrak{so}(3,1)$.

The two quadratic Casimir invariants of the Lie algebra AdS$_\omega$ are given by
\begin{gather}
{\cal C}=P_0^2-{\mathbf P}^2+\omega\big(J^2-{\mathbf K}^2\big),
\qquad
{\cal W}=-JP_0+K_1P_2-K_2P_1,
\label{bc}
\end{gather}
where here and in the following we denote ${\mathbf P}=(P_1,P_2)$, ${\mathbf P}^2=P_1^2+P_2^2$ and similarly for any other vector object
with two components.
Recall that $\cal C$ corresponds to the Killing--Cartan form and it is related to the energy of a~point particle, while
$\cal W$ is the Pauli--Lubanski vector.

It is well-known that parity $\Pi$, time-reversal $\Theta$ and their composition $\Pi\Theta$ def\/ined in~\cite{BLL}
\begin{alignat}{3}
& \Pi:
\  &&
(P_0, {\mathbf P},J, {\mathbf K})\to (P_0,-{\mathbf P},J,-{\mathbf K}),&
\nonumber
\\
& \Theta:
\ &&
(P_0, {\mathbf P},J, {\mathbf K})\to (-P_0, {\mathbf P},J,-{\mathbf K}),&
\nonumber
\\
& \Pi \Theta:
\ \ &&
(P_0, {\mathbf P},J, {\mathbf K})\to (-P_0,-{\mathbf P},J, {\mathbf K}),&
\label{inv}
\end{alignat}
are involutive automorphisms of AdS$_\omega$ which, together with the identity, def\/ine an Abelian group isomorphic to
$\mathbb Z_2
\times
\mathbb Z_2
\times
\mathbb Z_2$.
From this viewpoint, $\omega$ is a~graded contraction parameter related to the $\mathbb Z_2$-grading for
$\Pi\Theta$~\cite{Montigny} and gives rise to the following Cartan decomposition:
\begin{gather*}
\mbox{AdS$_\omega$} = \mathfrak{so}_\omega(2,2) ={\mathfrak{h}} \oplus {\mathfrak{p}},
\qquad
{\mathfrak{h}}=\text{span} \{J,{\mathbf K}\}=\mathfrak{so}(2,1),
\qquad
{\mathfrak{p}}=\text{span} \{P_0,{\mathbf P}\}.
\end{gather*}
It follows that the three classical (2+1)-dimensional Lorentzian symmetric homogeneous spacetimes with {\em constant
sectional curvature} $\omega$ are obtained as quotients ${\rm SO}_{\omega}(2,2) /{\rm SO}(2,1)$
where $H = {\rm SO}(2,1)$ and ${\rm SO}_{\omega}(2,2)$ are the Lie groups corresponding to ${\mathfrak{h}}$ and AdS$_\omega$,
respectively.
In the gravitational setting, the parameter $\omega$ is related to the {\em cosmological constant} $\Lambda$ via
$\omega=-\Lambda$.
More explicitly, we have the following description:
\begin{alignat*}{4}
&\omega>0,\ \Lambda<0:\ \mbox{AdS space}
\ \ \,  &&
  \omega=\Lambda=0:\ \mbox{Minkowski space}
\ \ \, &&
\omega<0,\ \Lambda>0:\ \mbox{dS space} &
\\
& {\mathbf {AdS}}^{2+1}={\rm SO}(2,2)/{\rm SO}(2,1)
\ \  \, &&
{\mathbf{M}}^{2+1}={\rm ISO}(2,1)/{\rm SO}(2,1)
\ \ \,  & &
{\mathbf {dS}}^{2+1}= {\rm SO}(3,1)/{\rm SO}(2,1) &
\end{alignat*}
Here, ${\rm SO}(2,1)$ is the Lorentz group in three dimensions whose Lie algebra $\mathfrak{so}(2,1)=\text{span}\{J,{\mathbf K}\}$
is spanned by the generators of boosts and spatial rotations, and the momenta $P_0$, ${\mathbf P}$ span the tangent space at the origin.
The curvature $\omega$ can also be written as $\omega=\pm 1/R^2$, where~$R$ is radius of the AdS and dS spaces, and the
limit $R\to \infty$ corresponds to their contraction to the Minkowski space.

\subsection{Vector model: ambient and geodesic parallel coordinates}

The action of the isometry groups ${\rm SO}_{\omega}(2,2)$ on their (2+1)-dimensional homogeneous spaces is nonlinear.
However, this problem can be circumvented by considering the vector representation of AdS$_\omega$ which makes use of an
ambient space with an ``extra'' dimension (called $s_3$ below).
This leads to the {\em vector representation} of AdS$_\omega$, in which the basis $\{P_0, {\mathbf P}, J, {\mathbf K}\}$ is represented
by the following $4\times4$ real matrices~\cite{CK3}:
\begin{alignat*}{4}
& P_0= \left(
\begin{matrix}
0&-\omega&0&0\\ 1&0&0&0\\ 0&0&0&0\\ 0&0&0&0
\end{matrix}
\right) ,
\qquad&&
P_1= \left(
\begin{matrix}
0&0&\omega&0\\ 0&0&0&0\\ 1&0&0&0\\ 0&0&0&0
\end{matrix}
\right) ,
\qquad&&
P_2= \left(
\begin{matrix}
0&0&0&\omega\\ 0&0&0&0\\ 0&0&0&0\\ 1&0&0&0
\end{matrix}
\right) ,&
\\
& J= \left(
\begin{matrix}
0&0&0&0\\ 0&0&0&0\\ 0&0&0&-1\\ 0&0&1&0
\end{matrix}
\right) ,
\qquad&&
K_1= \left(
\begin{matrix}
0&0&0&0\\ 0&0&1&0\\ 0&1&0&0\\ 0&0&0&0
\end{matrix}
\right) ,
\qquad&&
K_2= \left(
\begin{matrix}
0&0&0&0
\\
0&0&0&1
\\
0&0&0&0
\\
0&1&0&0
\end{matrix}
\right).&
\end{alignat*}
The corresponding one-parameter subgroups of ${\rm SO}_{\omega}(2,2)$ are obtained by exponentiation:
\begin{alignat}{3}
& {\rm e}^{x_0 P_0}= \left(
\begin{matrix}
\cos \eta x_0&-\eta \sin \eta x_0&0&0\\ \frac 1 \eta \sin \eta x_0&\cos \eta x_0&0&0\\ 0&0&1&0\\ 0&0&0&1
\end{matrix}
\right) ,
\quad &&
{\rm e}^{\theta J}= \left(
\begin{matrix}
1&0&0&0\\ 0&1&0&0\\ 0&0&\cos\theta&-\sin\theta\\ 0&0&\sin\theta&\cos\theta
\end{matrix}
\right) , &
\nonumber
\\
& {\rm e}^{x_1 P_1}= \left(
\begin{matrix}
\cosh \eta x_1&0&\eta \sinh \eta x_1&0\\ 0&1&0&0\\ \frac 1 \eta \sinh \eta x_1&0&\cosh \eta x_1&0\\ 0&0&0&1
\end{matrix}
\right) ,
\quad &&
{\rm e}^{\xi_1 K_1}= \left(
\begin{matrix}
1&0&0&0\\ 0&\cosh\xi_1&\sinh\xi_1&0\\ 0&\sinh\xi_1&\cosh\xi_1&0\\ 0&0&0&1
\end{matrix}
\right) ,&
\nonumber
\\
& {\rm e}^{x_2 P_2}= \left(
\begin{matrix}
\cosh \eta x_2&0&0&\eta \sinh \eta x_2\\ 0&1&0&0\\ 0&0&1&0\\ \frac 1 \eta \sinh \eta x_2&0&0&\cosh \eta x_2
\end{matrix}
\right) ,
\quad \ \;&&
{\rm e}^{\xi_2 K_2}= \left(
\begin{matrix}
1&0&0&0\\ 0&\cosh\xi_2&0&\sinh\xi_2\\ 0&0&1&0\\ 0&\sinh\xi_2&0&\cosh\xi_2
\end{matrix}
\right) ,\label{be}&
\end{alignat}
where hereafter the parameter $\eta$ is related to the curvature by $\omega=\eta^2=-\Lambda$.
This means that $\eta$ is either a~real number ($\eta=1/R$) for ${\bf AdS}^{2+1}$
or a~purely imaginary one ($\eta={\rm i}/R$) for ${\bf dS}^{2+1}$.

The matrix representation of the isometry groups ${\rm SO}_{\omega}(2,2)$ and their Lie algebras AdS$_\omega$ can be
characterised in terms of the bilinear form represented by the matrix
\begin{gather}
\mathbf I_{\omega}={\rm diag}(+1,\omega,-\omega,-\omega),
\label{bf}
\end{gather}
which identif\/ies them with the isometry groups of the four-dimensional linear space $(\mathbb R^4, \mathbf I_{\omega})$
with {\em ambient} or {\em Weierstrass} coordinates $(s_3,s_0,s_1,s_2)$:
\begin{gather*}
{\rm SO}_{\omega}(2,2)=\big\{G\in {\rm GL}(4,\mathbb R):\; G^T \mathbf I_{\omega} G=\mathbf I_{\omega} \big\},
\\
\mbox{AdS$_\omega$}=\big\{Y\in {\rm Mat}(4,\mathbb R):\; Y^T \mathbf I_{\omega}+\mathbf I_{\omega} Y=0 \big\},
\end{gather*}
where $A^T$ denotes the transpose of~$A$.
The origin of the ambient space has ambient coordinates $O =(1,0,0,0)$ and is invariant
under the Lorentz subgroup ${\rm SO} (2,1)\subset {\rm SO}_\omega(2,2)$ given by~\eqref{be}.
The orbit passing through~$O$ corresponds to the (2+1)-dimensional homogeneous spacetime which is contained in the
pseudosphere
\begin{gather*}
\Sigma_\omega\equiv s_3^2+\omega\big(s_0^2-{\mathbf s}^2\big)=1,
\end{gather*}
determined by $\mathbf I_{\omega}$~\eqref{bf}.
Note that in the limit $\omega\to 0$ ($R\to \infty$), which corresponds to the contraction to Minkowski space, the
pseudosphere $\Sigma_\omega$ gives rise to two hyperplanes, which are characterised by the condition $s_3=\pm 1$ in
Cartesian coordinates $(s_0,{\mathbf s})$.
From now on, we will identify the Minkowski space with the hyperplane given by $s_3=+1$.

The metric on the homogeneous spacetime is obtained from the f\/lat ambient metric given by~$\mathbf I_{\omega}$ by
dividing by the curvature and restricting the resulting metric to the pseudosphere $\Sigma_\omega$:
\begin{gather}
{\rm d}\sigma^2 = \left.{\frac 1 \omega} \left({\rm d} s_3^2+\omega\big({\rm d}s_0^2-{\rm d}s_1^2-{\rm d}s_2^2\big)\right)
\right|_{\Sigma_{\omega}}
\nonumber
\\
\phantom{{\rm d}\sigma^2}
={\rm d}s_0^2-{\rm d}s_1^2-{\rm d}s_2^2+\omega \frac{(s_0 {\rm d}s_0-s_1{\rm d}s_1-s_2{\rm d}s_2)^2}{1-\omega(s_0^2-{\mathbf s}^2)}.
\label{bh}
\end{gather}
Now let us introduce three {\em intrinsic} spacetime coordinates that will be helpful in the sequel: these are the
so-called {\em geodesic parallel coordinates} $(x_0,x_1,x_2)$~\cite{conformes}, which can be regarded as
a~generalisation of the f\/lat Cartesian coordinates to non-vanishing curvature.
They are def\/ined in terms of the action of the one-parameter subgroups~\eqref{be} for $P_0$, ${\mathbf P}$ on the origin
$O=(1,0,0,0)$:
\begin{gather}
(s_3,s_0,{\mathbf s})(x_0,{\mathbf x})=\exp(x_0 P_0)\exp(x_1 P_1)\exp(x_2 P_2) O,
\nonumber
\\
 s_3=\cos \eta x_0 \cosh \eta x_1 \cosh \eta x_2,
\qquad
s_1=\frac {\sinh \eta x_1}\eta \cosh \eta x_2,
\nonumber
\\
 s_0=\frac {\sin \eta x_0}\eta \cosh \eta x_1 \cosh \eta x_2,
\qquad
s_2=\frac {\sinh \eta x_2} \eta.
\label{bi}
\end{gather}
The geometrical meaning of the coordinates $(x_0,{\mathbf x})$ that parametrise a~generic point~$Q$ in the spacetime
via~\eqref{bi} is as follows.
Let $l_0$ be a~time-like geodesic and $l_1$, $l_2$ two space-like geodesics such that these three basis geodesics are
orthogonal at~$O$.
Then $x_0$ is the geodesic distance from~$O$ up to a~point $Q_1$ measured along the time-like geodesic~$l_0$; $x_1$ is
the geodesic distance between~$Q_1$ and another point~$Q_2$ along a~space-like geodesic $l'_1$ orthogonal to~$l_0$
through~$Q_1$ and parallel to~$l_1$; and~$x_2$ is the geodesic distance between~$Q_2$ and~$Q$ along a~space-like
geodesic~$l'_2$ orthogonal to~$l'_1$ through $Q_2$ and parallel to $l_2$.

Recall that time-like geodesics (as $l_0$) are compact in ${\bf AdS}^{2+1}$ and non-compact in ${\bf dS}^{2+1}$, while
space-like ones (as $l_i$, $l'_i$; $i=1,2$) are compact in ${\bf dS}^{2+1}$ but non-compact in ${\bf AdS}^{2+1}$.
Thus the trigonometric functions depending on $x_0$ are circular in ${\bf AdS}^{2+1}$ ($\eta=1/R)$ and hyperbolic in
${\bf dS}^{2+1}$ ($\eta={\rm i}/R)$ and, conversely, those depending on $x_i$ are circular in ${\bf dS}^{2+1}$ but
hyperbolic in ${\bf AdS}^{2+1}$.
By inserting the parametrisation~\eqref{bi} into the metric~\eqref{bh} we obtain the corresponding expression in terms
of geodesic parallel coordinates:
\begin{gather*}
{\rm d}\sigma^2 =\cosh^2(\eta x_1) \cosh^2(\eta x_2){\rm d} x_0^2-\cosh^2(\eta x_2){\rm d} x_1^2-{\rm d} x_2^2.
\end{gather*}
For $\omega\in\{\pm 1,0\}$ this expression reduces to
\begin{alignat*}{3}
& {\bf AdS}^{2+1}
\
(\omega=1,\; \eta=1):
\quad &&
{\rm d}\sigma^2 =\cosh^2 x_1 \cosh^2 x_2 \, {\rm d} x_0^2-\cosh^2 x_2 \, {\rm d} x_1^2-{\rm d} x_2^2,  &
\\
& {\bf M}^{2+1}
\
(\omega=\eta=0):
\quad &&
{\rm d}\sigma^2 = {\rm d} s_0^2-{\rm d} s_1^2-{\rm d} s_2^2= {\rm d} x_0^2-{\rm d} x_1^2-{\rm d}x_2^2, &
\\
& {\bf dS}^{2+1}
\
(\omega=-1,\; \eta = {\rm i}):
\quad &&
{\rm d}\sigma^2 =\cos^2 x_1 \cos^2 x_2\, {\rm d} x_0^2-\cos^2 x_2\, {\rm d} x_1^2-{\rm d} x_2^2 .&
\end{alignat*}

\subsection{Invariant vector f\/ields}

Left- and right-invariant vector f\/ields, $Y^L$ and $Y^R$, of the group $\rm {SO}_\omega(2,2)$ can be described in terms
of the matrix representation~\eqref{be}.
For this, one parametrises the group elements $T\in \rm {SO}_\omega(2,2)$ in terms of the matrices~\eqref{be} as
\begin{gather*}
T=\exp(x_0 P_0)\exp(x_1 P_1)\exp(x_2 P_2) \exp(\xi_1 K_1)\exp(\xi_2 K_2) \exp(\theta J).
\end{gather*}
This yields a~matrix of the form
\begin{gather*}
T= \left(
\begin{matrix}
s_3&A_{31}&A_{32}&A_{33}
\\
s_0&B_{01}&B_{02}&B_{03}
\\
s_1&B_{11}&B_{12}&B_{13}
\\
s_2&B_{21}&B_{22}&B_{23}
\end{matrix}
\right) ,
\end{gather*}
where the entries $A_{\alpha\beta}$ and $B_{\mu\nu}$ depend on all the group coordinates
$(x_0,{\mathbf x},\boldsymbol{\xi},\theta)$ and on the parameter $\eta$.
In the limit $\eta\to 0$, these expressions reduce to the well-known matrix representation of the Poincar\'e group $\rm {ISO}(2,1)$
\begin{gather*}
\lim_{\eta \to 0} T= \left(
\begin{matrix}
1&0&0&0
\\
x_0&\Lambda_{01}&\Lambda_{02}&\Lambda_{03}
\\
x_1&\Lambda_{11}&\Lambda_{12}&\Lambda_{13}
\\
x_2&\Lambda_{21}&\Lambda_{22}&\Lambda_{23}
\end{matrix}
\right) ,
\end{gather*}
where the entries $\Lambda_{\mu\nu}$ parametrise an element of ${\rm SO}(2,1)$ and depend only on
$(\boldsymbol{\xi},\theta)$.
From the group action of ${\rm SO}_\omega(2,2)$ on itself via right- and left-multiplication, one obtains expressions
for the left- and right-invariant vector f\/ields in terms of the coordinates $(x_0,{\mathbf x},\boldsymbol{\xi},\theta)$, which
are displayed in Table~\ref{table1}.
We stress that the limit $\eta\to 0$ of these expressions is always well def\/ined and gives the left- and right-invariant
vector f\/ields on the Poincar\'e group ${\rm ISO}(2,1)$.

\begin{table}[t!]\centering
 \footnotesize   \caption{Left- and right-invariant vector f\/ields on the isometry groups of the
(2+1)-dimensional (anti-)de Sitter and Minkowski spaces in terms of the sectional curvature $\omega=\eta^2=-\Lambda$.}
\label{table1}
\begin{tabular}{|l|}
\hline
\tsep{8pt}$\displaystyle{
Y_{P_0}^L=\frac{\cosh\xi_1\cosh\xi_2}{\cosh \eta x_1\cosh \eta x_2} (\partial_{x_0}-\eta \sinh \eta
x_1\partial_{\xi_1})+\frac{\sinh\xi_1\cosh\xi_2}{\cosh \eta x_2} \partial_{x_1}+\sinh\xi_2 \partial_{x_2}-\eta\tanh
\eta x_2 \cosh\xi_2 \partial_{\xi_2}
}$
\\
$\displaystyle{
Y_{P_1}^L= \left(\frac{\sinh\xi_1 \cos\theta+\cosh\xi_1\sinh\xi_2 \sin\theta}{\cosh \eta
x_1\cosh \eta x_2}\right)(\partial_{x_0}-\eta \sinh \eta x_1\partial_{\xi_1})+\left(\frac{\cosh\xi_1\cos\theta
+\sinh\xi_1\sinh\xi_2\sin\theta}{\cosh \eta x_2}\right)\partial_{x_1}
}$
\\
$\displaystyle{
\qquad
{} +\cosh \xi_2\sin\theta  \partial_{x_2}-\eta \tanh \eta x_2 \left(\tanh\xi_2 \cos\theta \partial_{\xi_1}+\sinh\xi_2
\sin\theta \partial_{\xi_2}-\frac{\cos\theta}{\cosh\xi_2} \partial_{\theta} \right)
}$
\\
$\displaystyle{Y_{P_2}^L= \left(\frac{\cosh\xi_1\sinh\xi_2\cos\theta-\sinh\xi_1\sin\theta}{\cosh \eta
x_1\cosh \eta x_2}\right)(\partial_{x_0}-\eta \sinh \eta x_1\partial_{\xi_1})
+\left(\frac{\sinh\xi_1\sinh\xi_2\cos\theta-\cosh\xi_1 \sin\theta}{\cosh \eta x_2}\right)\partial_{x_1}
}$
\\
$\displaystyle{
\qquad
{}+\cosh \xi_2\cos\theta \partial_{x_2}+\eta \tanh \eta x_2 \left(\tanh\xi_2 \sin\theta \partial_{\xi_1}-\sinh\xi_2
\cos\theta \partial_{\xi_2}- \frac{\sin\theta}{\cosh\xi_2} \partial_{\theta} \right)
}$
\\
$\displaystyle{
Y_{K_1}^L= \frac{\cos\theta}{\cosh \xi_2}  \partial_{\xi_1}+\sin\theta \partial_{\xi_2}+
\tanh \xi_2\cos\theta \partial_{\theta}
}$
\\
$\displaystyle{
Y_{K_2}^L=-\frac{\sin\theta}{\cosh \xi_2}  \partial_{\xi_1}+\cos\theta \partial_{\xi_2}-
\tanh \xi_2\sin\theta  \partial_{\theta}
}$
\\
$\displaystyle{
Y_{J}^L= \partial_{\theta}
}$\bsep{2pt}
\\
\hline
\tsep{5pt}$\displaystyle{
Y_{P_0}^R= \partial_{x_0}
}$
\\
$\displaystyle{
Y_{P_1}^R=-\sin \eta x_0\tanh \eta x_1  \partial_{x_0}+\cos \eta x_0 \partial_{x_1}-\eta \frac{\sin \eta x_0}{\cosh \eta x_1}\partial_{\xi_1}
}$
\\ $\displaystyle{
Y_{P_2}^R=-\frac{\sin \eta x_0\tanh \eta x_2}{\cosh \eta x_1}\left(\partial_{x_0}-\eta\sinh\eta x_1   \partial_{\xi_1} \right)
-\cos \eta x_0\sinh \eta x_1\tanh \eta x_2 \partial_{x_1}+\cos \eta x_0\cosh \eta x_1 \partial_{x_2}
}$
\\
$\displaystyle{
\qquad
{} +{\eta} \left(\frac{\cos \eta x_0\sinh \eta x_1\sinh \xi_1-\sin \eta x_0 \cosh \xi_1}{\cosh \eta x_2} \right)
\partial_{\xi_2}
}$
\\
$\displaystyle{
\qquad
{} +{\eta}\left(\frac {\cos \eta x_0\sinh \eta x_1\cosh \xi_1-\sin \eta x_0 \sinh \xi_1} {\cosh \eta x_2 \cosh
\xi_2}\right) \left(\partial_{\theta}-\sinh\xi_2 \partial_{\xi_1} \right)
}$
\\
$\displaystyle{
Y_{K_1}^R= \frac{\cos \eta x_0 \tanh \eta x_1}{\eta}\partial_{x_0}+\frac{\sin \eta
x_0}{\eta}\partial_{x_1}+\frac{\cos \eta x_0}{\cosh \eta x_1} \partial_{\xi_1}
}$
\\
$\displaystyle{
Y_{K_2}^R= \frac{\cos \eta x_0 \tanh \eta x_2}{\eta \cosh \eta x_1}\left(\partial_{x_0}-\eta\sinh \eta x_1   \partial_{\xi_1} \right)
-\frac{\sin \eta x_0 \sinh \eta x_1 \tanh \eta x_2}{\eta} \partial_{x_1}+\frac{\sin \eta x_0 \cosh \eta x_1}{\eta} \partial_{x_2}
}$
\\
$\displaystyle{
\qquad
{} +\left(\frac{\sin \eta x_0 \sinh\eta x_1 \sinh \xi_1+\cos \eta x_0\cosh \xi_1}{\cosh \eta x_2} \right)
\partial_{\xi_2}
}$
\\
$\displaystyle{
\qquad
{} +\left(\frac {\sin \eta x_0\sinh \eta x_1\cosh \xi_1+\cos \eta x_0 \sinh \xi_1} {\cosh \eta x_2 \cosh \xi_2} \right)
\left(\partial_{\theta}-\sinh\xi_2 \partial_{\xi_1} \right)
}$
\\
$\displaystyle{
Y_{J}^R=-\frac{\cosh \eta x_1 \tanh \eta x_2}{\eta} \partial_{x_1}+\frac{\sinh \eta
x_1}{\eta} \partial_{x_2}-\frac{\cosh \eta x_1}{\cosh \eta x_2}\left(\cosh\xi_1 \tanh\xi_2   \partial_{\xi_1}-
\sinh\xi_1   \partial_{\xi_2}-\frac{\cosh\xi_1}{\cosh\xi_2} \partial_{\theta} \right)
}$\bsep{8pt}
\\
\hline
\end{tabular}
\end{table}

\section[Twisted $\kappa$-AdS$_\omega$ algebra as a~Drinfel'd double]
{Twisted $\boldsymbol{\kappa}$-AdS$\boldsymbol{_\omega}$ algebra as a~Drinfel'd double}
\label{Section3}

We now consider the Lie bialgebra structures underlying the quantum deformations of the isometry groups ${\rm
SO}_\omega(2,2)$.
It is well-known that the Lie bialgebra structures underlying quantum deformations of semisimple Lie algebras are always
coboundary ones~\cite{BelDr} and hence characterised by classical~$r$-matrices.
This will be the case for all possible quantum deformations of the isometry groups of AdS and dS.
Moreover, it is well-known that all quantum deformations of the (2+1)-dimensional Poincar\'e algebra are also
coboundaries~\cite{ZakrCMP}.
These quantum deformations can therefore be classif\/ied by considering the classical~$r$-matrices of ${\rm AdS}_\omega$
and relating them via the cosmological constant.

The f\/irst steps towards such a~classif\/ication for the Lie algebras AdS$_\omega$ were recently presented
in~\cite{tallin}, where it became evident that there is a~plethora of possible quantum deformations.
Therefore, criteria to select the physically relevant cases are required.
In this respect, the Chern--Simons formulation of (2+1)-gravity can be helpful, since Poisson--Lie structures and Lie
bialgebra structures arise naturally in the description of its classical phase space.
The compatibility of a~given classical~$r$-matrix for ${\rm AdS}_\omega$ with the Chern--Simons formulation imposes
restrictions on the possible~$r$-matrices~\cite{cm1, cm2}.
We have recently shown in~\cite{BHMcqg} that these constraints are always fulf\/illed if the classical~$r$-matrix that
def\/ines the deformation is the canonical~$r$-matrix of certain Drinfel'd double structures of the Lie algebra ${\rm AdS}_\omega$.
In this section we show that the {\em twisted} $\kappa$-deformation is one of these compatible structures and thus
appears to be the appropriate generalisation of the $\kappa$-deformation in the context of (2+1)-gravity.

\subsection{Drinfel'd double Lie algebras}

A $2d$-dimensional Lie algebra $\mathfrak{a}$ has the structure of a~Drinfel'd double (DD)~\cite{Drinfelda} (see
also~\cite{Bais, BMS,BatistaMajid}) if there exists a~basis $\{Y_1,\dots,Y_d,y^1,\dots,y^d \}$ of $\mathfrak{a}$ in
which the Lie bracket takes the form
\begin{gather}
[Y_i,Y_j]= c^k_{ij}Y_k,
\qquad
[y^i,y^j]= f^{ij}_k y^k,
\qquad
[y^i,Y_j]= c^i_{jk}y^k-f^{ik}_j Y_k.
\label{agd}
\end{gather}
Note that this implies that $\{Y_1,\dots,Y_d\}$ and $\{y^1,\dots,y^d \}$ span two Lie subalgebras with structure
constants $c^k_{ij}$ and $f^{ij}_k$, respectively.
From the form of the ``crossed'' brackets $[y^i,Y_j]$, it follows that there is an Ad-invariant quadratic form on
$\mathfrak{a}$ given by
\begin{gather}\label{ages}
\langle Y_i,Y_j\rangle= 0,
\qquad
\langle y^i,y^j\rangle=0,
\qquad
\langle y^i,Y_j\rangle= \delta^i_j
\qquad
\forall\, i,j,
\end{gather}
and a~quadratic Casimir for $\mathfrak{a}$ is given by
\begin{gather}
C=\frac12\sum\limits_{i=1}^d{\big(y^i Y_i+Y_i y^i\big)}.
\label{cascas}
\end{gather}
A Lie algebra $\mathfrak{a}$ with a~DD structure can therefore be regarded as a~pair of Lie algebras, $\mathfrak{g}$
with basis $\{Y_1,\dots,Y_d\}$ and $\mathfrak{g} ^*$ with basis $\{y^1,\dots,y^d\}$, together with a~specif\/ic set of
crossed commutation rules~\eqref{agd} that ensures the existence of the Ad-invariant symmetric bilinear
form~\eqref{ages}.
We shall refer to Lie algebras with a~DD structure as {\em DD Lie algebras} in the following.

A DD Lie algebra $D ({\mathfrak{g}})\equiv \mathfrak{a}$ is always endowed with a~(quasi-triangular) Lie bialgebra
structure $(D ({\mathfrak{g}}),\delta_{D})$ that is generated by the canonical classical~$r$-matrix
\begin{gather}
r=\sum\limits_{i=1}^d{y^i\otimes Y_i},
\label{rcanon}
\end{gather}
through the coboundary relation
\begin{gather}
\delta_{D}(X_i)=[X_i \otimes 1+1\otimes X_i, r]
\qquad
\forall\, X_i\in D ({\mathfrak{g}}).
\label{rcanon2}
\end{gather}
Thus, the cocommutator $\delta_{D}$ takes the form
\begin{gather*}
\delta_{D}\big(y^k\big)=c^k_{ij} y^i\otimes y^j,
\qquad
\delta_{D}(Y_n)=-f_n^{lm} Y_l\otimes Y_m.
\end{gather*}
Note that the cocommutator $\delta_D$ only depends on the skew-symmetric component of the~$r$-matrix~\eqref{rcanon},
namely
\begin{gather}
r'=\frac12 \sum\limits_{i=1}^d{y^i\wedge Y_i},
\label{rmat}
\end{gather}
while the symmetric component of the~$r$-matrix def\/ines a~canonical quadratic Casimir element of $D(\mathfrak g)$ in the
form~\eqref{cascas}.
This implies that the associated element of $D(\mathfrak g)\otimes D(\mathfrak g)$ given by
\begin{gather*}
\Omega=r-r'=\frac12\sum\limits_{i=1}^d{\big(y^i\otimes Y_i+Y_i\otimes y^i\big)}
\end{gather*}
is invariant under the action of $D(\mathfrak g)$
\begin{gather*}
[Y_i \otimes 1+1\otimes Y_i,\Omega]=0
\qquad
\forall\, Y_i\in D(\mathfrak g).
\end{gather*}

To summarise, if a~Lie algebra $\mathfrak{a}$ has a~DD structure~\eqref{agd}, then this implies that
$(\mathfrak{a},\delta_{D})$ is a~Lie bialgebra with canonical~$r$-matrix~\eqref{rcanon}.
Therefore, there exists a~quantum algebra $(U_z(\mathfrak{a}),\Delta_z)$ whose f\/irst-order coproduct is given by
$\delta_{D}$~\eqref{rcanon2}, and this quantum deformation can be viewed as the quantum symmetry corresponding to the
given DD structure for $\mathfrak{a}$.

\subsection{The AdS Drinfel'd double}

As shown in~\cite{Witten1}, the Lie algebras $\mathfrak{so}(3,1)$, $\mathfrak{iso}(2,1)=\mathfrak{so}(2,1)\ltimes
\mathbb R^3$ and $\mathfrak{so}(2,2)$ of the isometry groups of Lorentzian (2+1)-gravity can be described in terms of
a~common basis in which the cosmological constant $\Lambda$ plays the role of a~structure constant.
In terms of the generators $T_a$ $(a=0,1,2)$ of translations and the generators $J_a$ $(a=0,1,2)$ of Lorentz
transformations, the Lie bracket then takes the form
\begin{alignat}{4}
& [J_0,J_1]=J_2,
\qquad &&
[J_0,J_2]=-J_1,
\qquad &&
[J_1, J_2]=- J_0,&\nonumber
\\
& [J_0,T_0]=0,
\qquad &&
[J_0,T_1]= T_2,
\qquad &&
[J_0, T_2]=-T_1, &\nonumber
\\
& [J_1,T_0]=-T_2,
\qquad &&
[J_1,T_1]=0,
\qquad &&
[J_1, T_2]=- T_0, &\label{jj}
\\
& [J_2,T_0]=T_1,
\qquad &&
[J_2,T_1]=  T_0,
\qquad &&
[J_2, T_2]=0,&\nonumber
\\
& [T_0,T_1]=-\Lambda   J_2,
\qquad &&
[T_0, T_2]= \Lambda  J_1,
\qquad &&
[T_1,T_2]= \Lambda   J_0.&\nonumber
\end{alignat}
As these are six-dimensional real Lie algebras, they can carry DD structures.
In that case, the DD structure provides a~canonical~$r$-matrix and, therefore, an associated quantum deformation
compatible with the constraints imposed by (2+1)-gravity.
The compatible DD structures on these Lie algebras were investigated in~\cite{BHMcqg}.
The twisted $\kappa$-AdS~$r$-matrix arises as the case F in the classif\/ication of admissible~$r$-matrices, which
corresponds to the DD $(6_0 | 5.iii | \lambda)$ in~\cite{Snobl} and case (11) in~\cite{gomez}.
This DD depends on an essential deformation parameter $\eta\neq 0$ and is explicitly given by
\begin{alignat*}{4}
& [Y_0,Y_1]=-Y_2, \qquad && [Y_0,Y_2]=-Y_1, \qquad && [Y_1,Y_2]=0, &
\\
& [y^0,y^1]=\eta y^1, \qquad && [y^0,y^2]=\eta y^2, \qquad && [y^1,y^2]=0,&
\end{alignat*}
together with the crossed relations
\begin{alignat*}{4}
& [y^0,Y_0]=0, \qquad && [y^0,Y_1]=-\eta Y_1, \qquad && [y^0,Y_2]=-\eta Y_2, &
\\
& [y^1,Y_0]=-y^2, \qquad && [y^1,Y_1]=\eta Y_0, \qquad && [y^1,Y_2]=y^0, &
\\
& [y^2,Y_0]=-y^1, \qquad && [y^2,Y_1]=y^0, \qquad && [y^2,Y_2]=\eta Y_0. &
\end{alignat*}

The change of basis that transforms these expressions into~\eqref{jj} is
\begin{alignat}{4}
&  J_0=\frac{1}{\sqrt{2\eta}}\big(Y_2-y^1\big), \qquad && J_1=\frac{1}{\sqrt{2\eta}}\big(Y_2+y^1\big), \qquad &&
J_2=-\frac 1 \eta y^0, &
\nonumber
\\
& T_0=\sqrt{\frac{\eta}{{2}}}\big(Y_1-y^2\big), \qquad && T_1=\sqrt{\frac{\eta}{{2}}}\big(Y_1+y^2\big),
\qquad && T_2=-\eta Y_0. &
\label{csbasis7}
\end{alignat}
Hence we obtain the AdS Lie algebra $\mathfrak{so}(2,2)$ with negative cosmological constant $\Lambda=-\eta^2$, which is
directly related to the deformation parameter $\eta$.
The canonical pairing~\eqref{ages} takes the form
\begin{gather}
\langle J_a,T_b\rangle=g_{ab},
\qquad
\langle J_a,J_b\rangle=\langle T_a,T_b\rangle=0,
\label{zzjj}
\end{gather}
where $g=\diag(-1,1,1)$ denotes the Minkowski metric in three dimensions.
We stress that~\eqref{zzjj} was shown in~\cite{Witten1} to be the appropriate pairing for the Chern--Simons formulation
of (2+1)-gravity, while other choices of pairing lead to a~dif\/ferent symplectic structure.

By inserting the inverse of~\eqref{csbasis7} into the canonical classical~$r$-matrix~\eqref{rcanon} and by using the
Casimir operator~\eqref{cascas} in order to get a~fully skew-symmetric expression as in~\eqref{rmat}, we f\/ind that the
AdS deformation induced by this DD structure is generated by
\begin{gather}
r'_{\rm F} =\tfrac{1}{2} (J_1\wedge T_0-J_0 \wedge T_1+J_2\wedge T_2).
\label{zm}
\end{gather}

Since the Lie algebra elements $J_a$ are the generators of the Lorentz group and the generators~$T_a$ generate the
(non-commutative) AdS translation sector $(a=0,1,2)$, we conclude that the classical~$r$-matrix~\eqref{zm} is just
a~superposition of the standard deformation of~$\mathfrak{so}(2,2)$ \cite{CK3} generated by $(J_1\wedge T_0-J_0 \wedge
T_1)$ and a~Reshetikhin twist generated by~$J_2\wedge T_2$ (note that~$J_2$ and~$T_2$ commute).
Therefore, we have obtained a~twisted $\kappa$-AdS algebra which is a~DD structure.

\subsection{The dS Drinfel'd double}

The analogous DD deformation of $\mathfrak{so}(3,1)$ is given by case (9) in~\cite{gomez} and case $(7_0 | 5.ii |
\lambda)$ in~\cite{Snobl}, and corresponds to case C in~\cite{BHMcqg}.
It depends again on one essential deformation parameter $\eta\neq 0$, and the Lie bracket is given by
\begin{alignat*}{4}
& [Y_0,Y_1]= Y_2, \qquad && [Y_0,Y_2]=-Y_1, \qquad && [Y_1,Y_2]=0, &
\\
& [y^0,y^1]=-\eta y^1, \qquad && [y^0,y^2]=-\eta y^2, \qquad && [y^1,y^2]=0, &
\end{alignat*}
with crossed commutators
\begin{alignat*}{4}
& [y^0,Y_0]=0, \qquad && [y^0,Y_1]=\eta Y_1, \qquad && [y^0,Y_2]=\eta Y_2, &
\\
& [y^1,Y_0]=-y^2, \qquad && [y^1,Y_1]=-\eta Y_0, \qquad && [y^1,Y_2]=y^0, &
\\
& [y^2,Y_0]=y^1, \qquad && [y^2,Y_1]=-y^0, \qquad && [y^2,Y_2]=-\eta Y_0. &
\end{alignat*}
To obtain a~Lie algebra isomorphism between this DD Lie algebra and the isometry algebra~\eqref{jj} of dS space, we
consider the change of basis
\begin{alignat*}{4}
& J_0=\frac{1}{\sqrt{2\eta}}\big(Y_1-y^2\big), \qquad && J_1=\frac{1}{\sqrt{2\eta}}\big(Y_1+y^2\big), \qquad && J_2= \frac 1 {\eta}y^0, &
\\
& T_0=\sqrt{\frac{\eta}{{2}}}\big(Y_2-y^1\big), \qquad && T_1=\sqrt{\frac{\eta}{{2}}}\big(Y_2+y^1\big), \qquad && T_2=\eta Y_0. &
\end{alignat*}
This yields the Lie algebra $\mathfrak{so}(3,1)$ with bracket~\eqref{jj} and positive cosmological constant
$\Lambda=\eta^2$, together with the same canonical pairing~\eqref{zzjj}.
Using again the Casimir operator~\eqref{cascas}, one obtains the skew-symmetric classical~$r$-matrix from the canonical
one~\eqref{rcanon}
\begin{gather}
r'_{\rm C} = \tfrac{1}{2} (J_1\wedge T_0-J_0 \wedge T_1+J_2\wedge T_2) \equiv r'_{\rm F}. \label{xp}
\end{gather}
As the~$r$-matrix $r'_{\rm F}= r'_{\rm C}$ does not depend on the deformation parameter $\eta$ and, consequently, is
independent of the cosmological constant, it is a~{\em common} classical~$r$-matrix for the {\em three} DD structures
$\mathfrak{so}(2,2)$, $\mathfrak{so}(3,1)$ and $\mathfrak{iso}(2,1)$ on the (2+1)-dimensional Lorentzian spacetimes
${\bf AdS}^{2+1}$, ${\bf dS}^{2+1}$ and ${\bf M}^{2+1}$, the latter being obtained from the limit $\eta\to 0$.

\subsection[The AdS$_\omega$ Drinfel'd double in the kinematical basis]{The AdS$\boldsymbol{_\omega}$ Drinfel'd double in the kinematical basis}

We now analyse the DD structures of the Lie algebra AdS$_\omega$ in the kinematical basis $\{P_0,{\mathbf P}, J, {\mathbf K}\}$ with
commutation relations~\eqref{ba}.
Recall that AdS$_\omega$ gives a~unif\/ied description of the three Lie algebras $\mathfrak{so}(2,2)$,
$\mathfrak{so}(3,1)$ and $\mathfrak{iso}(2,1)$, which are, accordingly, parametrised by the constant sectional curvature
$\omega$ of their corresponding homogeneous spacetimes ${\bf AdS}^{2+1}$, ${\bf dS}^{2+1}$ and ${\bf M}^{2+1}$.
In particular, we show that the classical~$r$-matrix~\eqref{xp} def\/ines {\em two} dif\/ferent quantum DD deformations
that, according to~\cite{CK4, CK3}, we shall call {\em space-like} and {\em time-like} deformations.

\subsubsection[The space-like~$r$-matrix]{The space-like~$\boldsymbol{r}$-matrix}\label{Section3.4.1}

It is easy to see that the change of basis
\begin{gather}
P_0=T_0,
\qquad
P_1=T_1,
\qquad
P_2=T_2,
\qquad
K_1=J_2,
\qquad
K_2=-J_1,
\qquad
J=J_0,
\label{xq}
\end{gather}
transforms~\eqref{jj} into~\eqref{ba} provided that $\omega=-\Lambda$.
Consequently, the deformation para\-me\-ter~$\eta$ above coincides with the one introduced in Section~\ref{Section2} in the form
$\omega=\eta^2$, assuming that $\eta$ is a~real number for AdS but a~purely imaginary one for dS.
Using~\eqref{xq}, one f\/inds that the classical~$r$-matrix~\eqref{xp} is given by
\begin{gather}
r'_{\rm F} =r'_{\rm C} =\tfrac{1}{2} (P_0\wedge K_2+P_1 \wedge J)+\tfrac{1}{2}  K_1\wedge P_2.
\label{zmll}
\end{gather}
To construct the associated quantum deformation, we scale this~$r$-matrix by a~quantum deformation parameter~$z$
(dif\/ferent from $\eta$) as
\begin{gather}
r^{\rm space}_z =2 z r'_{\rm F}=z (P_0\wedge K_2+P_1 \wedge J)+z K_1\wedge P_2,
\label{zmllb}
\end{gather}
This allows us to relate~\eqref{zmllb} to the results in~\cite{CK3}.
It becomes apparent that the f\/irst term in~\eqref{zmllb} is just a~{\em space-like}~$r$-matrix of type (a) for
AdS$_\omega$, while the second one is a~twist.
From~\eqref{rcanon2} it follows that the primitive (non-deformed) generators are $P_2$ and $K_1$.
Note that for the space-like~$r$-matrix the physical dimension of~$z$ is determined by the generator $P_2$ of spatial
translations $[z]=[P_2]^{-1}$, which implies that~$z$ has the dimension of a~{\em length}.
It is related to the usual deformation parameters $\kappa$ and~$q$ by
\begin{gather}
z=1/\kappa,
\qquad
q={\rm e}^z.
\label{kk}
\end{gather}
The classical limit of the quantum deformation, therefore, corresponds to $z\to 0$ ($\kappa\to \infty$, $q\to 1$).
Accordingly, we shall call~\eqref{zmllb} the {\em twisted $\kappa$-space-like~$r$-matrix}.

Moreover, the Lie algebra isomorphism $\mathfrak{so}(2,2)\simeq \mathfrak{sl}(2,\mathbb{R})\oplus
\mathfrak{sl}(2,\mathbb{R})$ suggests that a~(real) change of basis should exist between the AdS$_\omega$
basis~\eqref{ba} and that of two copies of $\mathfrak{sl}(2,\mathbb{R})$ when $\omega=\eta^2>0$ ($\eta\in\mathbb{R})$.
A direct computation shows that such an isomorphism is given by
\begin{gather*}
 P_0=\frac{\sqrt{\eta}}{2\sqrt{2}} \left(J_+^1+J_+^2-2 \eta\big(J_-^1+J_-^2\big)\right) ,
\qquad
J=\frac{1}{2\sqrt{2\eta}} \left(J_+^1-J_+^2+2 \eta\big({-}J_-^1+J_-^2\big)\right) ,
\\
P_1=\frac{\sqrt{\eta}}{2\sqrt{2}}
\left(J_+^1+J_+^2+2 \eta\big(J_-^1+J_-^2\big)\right) ,
\qquad
K_1=\frac{1}{2}\big(J_3^1+J_3^2\big),
\\
P_2=\frac{\eta}{2}\big(J_3^1-J_3^2\big),
\qquad
K_2=\frac{1}{2\sqrt{2\eta}}\left(-J_+^1+J_+^2+2 \eta\big({-}J_-^1+J_-^2\big)\right) ,
\end{gather*}
where
\begin{gather*}
[J_3^l,J_\pm^l] =\pm 2 J_\pm^l,
\qquad
[J_+^l,J_-^l] = J_3^l,
\qquad
l=1,2,
\end{gather*}
and all other Lie brackets vanish.
In this basis, the classical~$r$-matrix~\eqref{zmll} reads
\begin{gather}
r_F'=\frac{\eta}{2}\big({-}J_+^1\wedge J_-^1+J_+^2\wedge J_-^2\big)-\frac{\eta}{4} J_3^1\wedge J_3^2.
\label{rslsl}
\end{gather}
It becomes apparent that it is the superposition of two standard (Drinfel'd--Jimbo~\cite{Drinfelda,Jimbo}) deformations,
$J_+^l\wedge J_-^l$, on each of the two copies of $\mathfrak{sl}(2,\mathbb{R})$ but with {\em opposite sign}, and of
a~twist generated by $J_3^1\wedge J_3^2$.
This identif\/ication $U_z(\mathfrak{so}(2,2))\simeq U_z(\mathfrak{sl}(2,\mathbb{R}))\oplus
U_{-z}(\mathfrak{sl}(2,\mathbb{R}))$ was f\/irst introduced in~\cite{CGST1,CGST2} and further investigated
in~\cite{beyond}.

\subsubsection[The time-like~$r$-matrix]{The time-like~$\boldsymbol{r}$-matrix}

Alternatively, the classical~$r$-matrix~\eqref{xp} can be expressed as a~superposition of the {\em time-like}~$r$-matrix
for AdS$_\omega$~\cite{CK3} with a~twist.
This requires the complex change of basis
\begin{alignat}{4}
& P_0={\rm i} T_2, \qquad && P_1=-{\rm i} T_0, \qquad && P_2=-T_1, &\nonumber
\\
& K_1=J_1, \qquad && K_2=-{\rm i} J_0, \qquad && J=-{\rm i} J_2, &
\label{xxbb}
\end{alignat}
which again transforms~\eqref{jj} into~\eqref{ba} with $\omega=-\Lambda$.
The classical~$r$-matrix~\eqref{xp} then takes the form
\begin{gather*}
r'_{\rm F} =r'_{\rm C} =\tfrac{{\rm i}}{2} (K_1\wedge P_1+K_2 \wedge P_2)+\tfrac{1}{2} J\wedge P_0.
\end{gather*}
Introducing again the quantum deformation parameter~\eqref{kk} via a~rescaling we can write
\begin{gather}
r^{\rm time}_z=-2 {\rm i} z  r'_{\rm F} =z (K_1\wedge P_1+K_2 \wedge P_2)-{\rm i} z J\wedge P_0.
\label{zmllbb}
\end{gather}
In this case, the f\/irst term in~\eqref{zmllbb} is the {\em time-like}~$r$-matrix of type (b) for
AdS$_\omega$~\cite{CK3}, which coincides with the $\kappa$-Poincar\'e~$r$-matrix~\cite{LukNR, Maslanka} when
$\omega=\Lambda=\eta=0$, and the second one is the twist.
The primitive generators are now $P_0$ and~$J$ and the dimensions of~$z$ are given by those of the generator of time
translations $P_0$ as $[z]=[P_0]^{-1}$.
This implies that~$z$ has the dimension of a~{\em time}.
Accordingly, we shall call~\eqref{zmllbb} the {\em twisted $\kappa$-time-like~$r$-matrix}.

Similarly, we can also consider the isomorphism $\mathfrak{so}(2,2)\simeq \mathfrak{sl}(2,\mathbb{R})\oplus
\mathfrak{sl}(2,\mathbb{R})$, and apply the (complex) change of basis
\begin{gather*}
 P_0={\rm i}\frac{\eta}{2}\big(J_3^1-J_3^2\big),
\qquad
J=-\frac{{\rm i}}{2}\big(J_3^1+J_3^2\big),
\\
P_1=-\frac {\rm i} 2 \sqrt{\frac{\eta}{2}} \left(J_+^1+J_+^2-2 \eta\big(J_-^1+J_-^2\big)\right) ,
\qquad
K_1=\frac{1}{2\sqrt{2\eta}} \left(J_+^1-J_+^2+2 \eta\big(J_-^1-J_-^2\big)\right) ,
\\
P_2=-\frac {\rm i} 2
\sqrt{\frac{\eta}{2}} \left(J_+^1+J_+^2+2 \eta\big(J_-^1+J_-^2\big)\right) ,
\qquad
K_2=-\frac{{\rm 1}}{2\sqrt{2\eta}} \left(J_+^1-J_+^2-2 \eta\big(J_-^1-J_-^2\big)\right) ,
\end{gather*}
which gives rise to the same~$r$-matrix~\eqref{rslsl}.

\section[Lie bialgebra of the twisted $\kappa$-AdS$_\omega$ algebra]
{Lie bialgebra of the twisted $\boldsymbol{\kappa}$-AdS$\boldsymbol{_\omega}$ algebra}
\label{Section4}

So far, we have obtained a~common classical~$r$-matrix~\eqref{xp} from DD structures for the three Lie algebras that
form the family AdS$_\omega$.
Moreover, we have identif\/ied two physically distinct quantum deformations: the twisted $\kappa$-space-like deformation
and the twisted $\kappa$-time-like one.
As the latter (without the twist motivated by (2+1)-gravity) has been widely studied in the literature due to the role
of the deformation parameter $z=1/\kappa$ as a~fundamental time or energy scale possibly related to the Planck length,
we construct its full quantisation in the following.
Note, however, that in all the expressions to be presented in the sequel, it will always be possible to obtain the
corresponding $\kappa$-space-like counterparts by simply applying the map between both bases provided by~\eqref{xq}
and~\eqref{xxbb}, namely
\begin{alignat*}{5}
& {\mbox{time-like $\to$ space-like:}}  \quad && P_0\to {\rm i} P_2, \qquad && P_1\to-{\rm i} P_0, \qquad && P_2\to-P_1, &
\\
& && J\to-{\rm i} K_1, \qquad && K_1\to-K_2, \qquad && K_2 \to-{\rm i} J.&
\end{alignat*}
Moreover, in order to highlight the ef\/fect of $J\wedge P_0$ in the twisted $\kappa$-time-like deformation of
AdS$_\omega$ with classical~$r$-matrix~\eqref{zmllbb}, we shall consider the {\em two-parameter} classical~$r$-matrix
given~by
\begin{gather}
r=z(K_1\wedge P_1+K_2\wedge P_2)+\vartheta J\wedge P_0,
\label{ca}
\end{gather}
where $\vartheta$ is a~generic deformation parameter associated to the twist, that for $\vartheta=-{\rm i} z$ yields the
underlying DD structure.
We also recall that the~$r$-matrix~\eqref{ca} arises as a~particular case within the classif\/ication of deformations of
AdS$_\omega$ given in~\cite{tallin} for
\begin{gather*}
a_i=b_i=0,
\quad
i=1,2,3,4,
\qquad
a_5=b_5=-z,
\qquad
a_6=b_6=0,
\\
c_1=\vartheta,
\qquad
c_2= c_3=0,
\end{gather*}
which is a~solution of the modif\/ied classical Yang--Baxter equation with Schouten bracket
\begin{gather*}
[[r,r]]=-z^2 (P_0 \wedge P_1\wedge K_1+P_0\wedge P_2\wedge K_2+P_1\wedge P_2\wedge J)-z^2 \omega   K_1\wedge
K_2\wedge J.
\end{gather*}
The f\/irst term is just the Schouten bracket for the $\kappa$-Poincar\'e~$r$-matrix, while the second one includes the
ef\/fect of the curvature or cosmological constant in the AdS and dS cases with $\omega\ne 0$.
As expected, the twist does not af\/fect the Schouten bracket.

The Lie bialgebra $(\mbox{AdS$_\omega$},\delta)$ generated by~\eqref{ca} can be computed via~\eqref{rcanon2}, which
yields the cocommutator
\begin{gather}
\delta(P_0) = \delta(J)=0,
\nonumber
\\
\delta (P_1)= z (P_1\wedge P_0-\omega K_2\wedge J)+\vartheta (P_0\wedge P_2+\omega K_1\wedge J),
\nonumber
\\
\delta(P_2)= z (P_2\wedge P_0+\omega K_1 \wedge J)-\vartheta (P_0\wedge P_1-\omega K_2 \wedge J),
\nonumber
\\
\delta(K_1)= z (K_1 \wedge P_0+P_2 \wedge J)+\vartheta (P_0\wedge K_2-P_1\wedge J),
\nonumber
\\
\delta(K_2)= z (K_2 \wedge P_0-P_1\wedge J)-\vartheta (P_0\wedge K_1+P_2\wedge J).
\label{cc}
\end{gather}

Denoting by $\{x_0,{\mathbf x},\theta,\boldsymbol{\xi}\}$ the dual non-commutative coordinates of the generators $\{P_0, {\mathbf P}, J$,
${\mathbf K}\}$, respectively, one obtains from the cocommutators~\eqref{cc} the following dual Lie brackets between the
non-commutative spacetime coordinates
\begin{gather}
[\hat x_0, \hat x_1]=-z \hat x_1-\vartheta \hat x_2,
\qquad
[\hat x_0, \hat x_2]=-z \hat x_2+\vartheta \hat x_1,
\qquad
[\hat x_1, \hat x_2]=0,
\label{cd}
\end{gather}
as well as
\begin{gather}
[\hat\theta,\hat x_i]= z\epsilon_{ij}\hat \xi_j+\vartheta \hat\xi_i
\qquad
[\hat\theta,\hat \xi_i]=-\omega \left(z \epsilon_{ij} \hat x_j+\vartheta \hat x_i\right) ,
\qquad
[\hat\theta,\hat x_0]=0,
\nonumber
\\
[\hat x_0, \hat \xi_i]=-z \hat \xi_i-\vartheta \epsilon_{ij} \hat \xi_j,
\qquad
[\hat \xi_1, \hat \xi_2]=0,
\qquad
[\hat x_i,\hat \xi_j]=0,
\qquad
i,j=1,2.
\label{ce}
\end{gather}
Note that the expressions~\eqref{cd} are just the {\em first-order} of the non-commutative twisted $\kappa$-AdS$_\omega$
spacetimes, that will be constructed in the following sections.
Nevertheless, these expressions are suf\/f\/icient to analyse the role of both the cosmological constant $\Lambda=-\omega$
and the twist with quantum parameter $\vartheta$ and to compare them to the well-known $\kappa$-Minkowski spacetime
(see~\cite{RoxPLB, LukR,LukNR,Maslanka,kZakr} and references therein) that is given by
\begin{gather}
[\hat x_0, \hat x_1]=-z \hat x_1,
\qquad
[\hat x_0, \hat x_2]=-z \hat x_2,
\qquad
[\hat x_1, \hat x_2]=0,
\qquad
z=1/\kappa.
\label{cf}
\end{gather}
It was already shown in~\cite{Rox,tallin} that, at the f\/irst-order in the quantum coordinates, both non-commutative AdS
and dS spacetimes coincide with the $\kappa$-Minkowski one~\eqref{cf} and that (see~\cite{Rox}) the parameter
$\omega=\eta^2$ only enters the higher-order terms of the corresponding quantum deformation.

\section[The twisted $\kappa$-AdS$_\omega$ quantum algebra]{The twisted $\boldsymbol{\kappa}$-AdS$\boldsymbol{_\omega}$ quantum algebra}
\label{Section5}

In this section we construct the twisted $\kappa$-AdS$_\omega$ algebra with underlying Lie bialgebra given by~\eqref{ba} and~\eqref{cc}.
We f\/irst present this algebra in the so-called {\em symmetrical} kinematical basis~\cite{Rox,tallin}, in which the
corresponding algebra without the $\vartheta$-twist was obtained in~\cite{CK3}.
We then present the twisted $\kappa$-AdS$_\omega$ quantum deformation in a~{\em bicrossproduct-type} basis, which is
characterised by the fact that its limit $\omega\to 0$ gives rise to the $\kappa$-Poincar\'e algebra endowed with
a~proper bicrossproduct structure~\cite{majid,Majid:1994cy} (see also~\cite{bicross,azc, AP}).
Finally, we compare these results to quantum deformations investigated in the literature~\cite{Dasz1}.

\subsection{``Symmetrical'' basis}\label{Section5.1}

We start by recalling the expressions for the quantum $\kappa$-AdS$_\omega$ algebra in terms of the kinematical
basis~\eqref{ba}, as obtained in~\cite{CK3}.
The corresponding coproduct and compatible deformed commutation rules for the $\kappa$-AdS$_\omega$ algebra read
\begin{gather}
\Delta_z (P_0)=1\otimes P_0+P_0\otimes 1,
\qquad
\Delta_z(J)=1\otimes J+J\otimes 1,
\nonumber
\\
\Delta_z(P_i)={\rm e}^{-\frac z2 P_0}\cosh\left(\frac z2 \eta J\right)\otimes P_i+P_i \otimes {\rm e}^{\frac z2 P_0}\cosh\left(\frac z2 \eta J\right)
\nonumber
\\
\phantom{\Delta_z(P_i)=}
{}+\eta  {\rm e}^{-\frac z2 P_0}\sinh\left(\frac z2 \eta J\right)\otimes \epsilon_{ij} K_j-\eta  \epsilon_{ij} K_j \otimes {\rm
e}^{\frac z2 P_0}\sinh\left(\frac z2 \eta J\right),
\nonumber
\\
\Delta_z(K_i)={\rm e}^{-\frac z2 P_0}\cosh\left(\frac z2 \eta J\right)\otimes K_i+K_i \otimes {\rm e}^{\frac z2 P_0}\cosh\left(\frac z2 \eta J\right)
\nonumber
\\
\phantom{\Delta_z(K_i)=}
{} -{\rm e}^{-\frac z2 P_0}  \frac{\sinh(\frac z2 \eta J)}{\eta}\otimes \epsilon_{ij} P_j+\epsilon_{ij}
P_j \otimes {\rm e}^{\frac z2 P_0}   \frac{\sinh(\frac z2 \eta J)}{\eta},
\label{fa}
\\
[J,P_i]= \epsilon_{ij}P_j,
\qquad
[J,K_i]= \epsilon_{ij}K_j,
\qquad
[J,P_0]= 0,
\nonumber
\\
[P_i,K_j]=-\delta_{ij}\frac{\sinh (zP_0)}{z}\cosh(z\eta J),
\qquad
[P_0,K_i]=-P_i,
\qquad
[P_0,P_i]=\omega K_i,
\nonumber
\\
[P_1,P_2]=-\omega \cosh (zP_0)\frac{\sinh(z\eta J)}{z\eta},
\qquad
[K_1,K_2]=-\cosh (zP_0)\frac{\sinh(z\eta J)}{z\eta},
\label{fb}
\end{gather}
with $\omega=\eta^2=-\Lambda$.
The quantum deformation of the two Casimir invariants~\eqref{bc} reads
\begin{gather}
{\cal C}=4 \cos (z\eta) \left\{  \frac{\sinh^2(\frac z 2 P_0)}{z^2}   \cosh^2\left(\frac z 2\eta
J\right)+ \frac{\sinh^2(\frac z 2 \eta J)}{z^2}  \cosh^2\left(\frac z 2 P_0\right) \right\}\nonumber\\
\hphantom{{\cal C}=}{}
-\frac{\sin (z\eta)}{z\eta} \left({\mathbf P}^2+\omega {\mathbf K}^2 \right),
\nonumber
\\
 {\cal W}=-\cos (z\eta)\frac{\sinh(z\eta J)}{z\eta}  \frac{\sinh(z P_0)}{z}+\frac{\sin
(z\eta)}{z\eta} (K_1P_2-K_2P_1) .
\label{fc}
\end{gather}
With these results, we construct the twisted $(\vartheta,\kappa)$-quantum AdS$_\omega$ algebra with general twist
parameter $\vartheta$ by applying the well-known twisting procedure~\cite{Drinfeldb}.
This preserves the deformed commutation relations~\eqref{fb} and hence the Casimir invariants~\eqref{fc}.
The new coproduct $\Delta_{\vartheta,z}$ is obtained by twisting~\eqref{fa} with an element ${\cal F}_\vartheta\in
\kappa{\mbox {-AdS$_\omega$}} \otimes \kappa{\mbox {-AdS$_\omega$}}$ given by
\begin{gather}
\Delta_{\vartheta,z}(Y)={\cal F}_\vartheta\Delta_z(Y) {\cal F}_\vartheta^{-1}
\quad
\forall\,  Y\in \mbox{$\kappa$-AdS$_\omega$},
\qquad
\text{where}
\quad
{\cal F}_\vartheta=\exp(-\vartheta J \wedge P_0).
\label{twist}
\end{gather}
It satisf\/ies the so-called twisting co-cycle and normalisation conditions
\begin{gather*}
{\cal F}_{\vartheta,12}(\Delta_z \otimes \mathrm{id}){\cal F_\vartheta}={\cal F}_{\vartheta,23}(\mathrm{id}\otimes\Delta_z){\cal F}_\vartheta,
\qquad
(\epsilon\otimes\mathrm{id}){\cal F}_\vartheta=1=(\mathrm{id}\otimes\epsilon){\cal F}_\vartheta,
\end{gather*}
where \looseness=1 ${\cal F}_{\vartheta,12}={\cal F}_\vartheta\otimes\mathrm{id}$, ${\cal F}_{\vartheta,23}=\mathrm{id}\otimes{\cal
F}_\vartheta$ and $\epsilon$ is the co-unit map, $\epsilon(Y)=0$
$\forall \, Y\in \mbox{AdS$_\omega$}$.
The explicit form of the twisted coproduct is obtained through cumbersome computations and reads
\begin{gather*}
\Delta_{\vartheta,z}(P_0) = 1\otimes P_0+P_0\otimes 1,
\qquad
\Delta_{\vartheta,z}(J) = 1\otimes J+J\otimes 1,
\\
\Delta_{\vartheta,z}(P_i)
=\Delta_z (P_i)+{\rm e}^{-\frac{z}{2}P_0}\cosh(\tfrac{z}{2}\eta J)\left[\cos(\vartheta\eta J)\cos(\vartheta P_0)-1\right]\otimes P_i
\\
\phantom{\Delta_{\vartheta,z}(P_i)=}
{}+{\rm e}^{-\frac{z}{2}P_0}\cosh(\tfrac{z}{2}\eta J)\sin(\vartheta P_0)\cos(\vartheta\eta J)\otimes \epsilon_{ij}P_j
\\
\phantom{\Delta_{\vartheta,z}(P_i)=}
{}-\eta {\rm e}^{-\frac{z}{2}P_0}\cosh(\tfrac{z}{2}\eta J)\sin(\vartheta\eta J)\cos(\vartheta P_0)\otimes K_i
\\
\phantom{\Delta_{\vartheta,z}(P_i)=}
{}-\eta {\rm e}^{-\frac{z}{2}P_0}\cosh(\tfrac{z}{2}\eta J)\sin(\vartheta\eta J) \sin(\vartheta P_0)\otimes \epsilon_{ij}K_j
\\
\phantom{\Delta_{\vartheta,z}(P_i)=}
{}+P_i\otimes {\rm e}^{\frac{z}{2}P_0}\cosh(\tfrac{z}{2}\eta J)\left[\cos(\vartheta\eta J) \cos(\vartheta P_0)-1\right]
\\
\phantom{\Delta_{\vartheta,z}(P_i)=}
{}-\epsilon_{ij}P_j\otimes {\rm e}^{\frac{z}{2}P_0}\cosh(\tfrac{z}{2}\eta J) \sin(\vartheta P_0) \cos(\vartheta\eta J)
\\
\phantom{\Delta_{\vartheta,z}(P_i)=}
{}+\eta K_i\otimes {\rm e}^{\frac{z}{2}P_0}\cosh(\tfrac{z}{2}\eta J)\sin(\vartheta\eta J)\cos(\vartheta P_0)
\\
\phantom{\Delta_{\vartheta,z}(P_i)=}
{}-\eta \epsilon_{ij}K_j\otimes {\rm e}^{\frac{z}{2}P_0}\cosh(\tfrac{z}{2}\eta J)\sin(\vartheta\eta J)\sin(\vartheta P_0)
\\
\phantom{\Delta_{\vartheta,z}(P_i)=}
{}-{\rm e}^{-\frac{z}{2}P_0}\sinh(\tfrac{z}{2}\eta J)\sin(\vartheta\eta J)\sin(\vartheta P_0)\otimes P_i
\\
\phantom{\Delta_{\vartheta,z}(P_i)=}
{}+{\rm e}^{-\frac{z}{2}P_0}\sinh(\tfrac{z}{2}\eta J)\sin(\vartheta\eta J)\cos(\vartheta P_0)\otimes \epsilon_{ij}P_j
\\
\phantom{\Delta_{\vartheta,z}(P_i)=}
{}-\eta {\rm e}^{-\frac{z}{2}P_0}\sinh(\tfrac{z}{2}\eta J)\sin(\vartheta P_0)\cos(\vartheta\eta J)\otimes K_i
\\
\phantom{\Delta_{\vartheta,z}(P_i)=}
{}+\eta {\rm e}^{-\frac{z}{2}P_0}\sinh(\tfrac{z}{2}\eta J)[\cos(\vartheta\eta J)\cos(\vartheta P_0)-1]\otimes\epsilon_{ij}K_j
\\
\phantom{\Delta_{\vartheta,z}(P_i)=}
{}+P_i\otimes {\rm e}^{\frac{z}{2}P_0}\sinh(\tfrac{z}{2}\eta J)\sin(\vartheta\eta J)\sin(\vartheta P_0)
\\
\phantom{\Delta_{\vartheta,z}(P_i)=}
{}+\epsilon_{ij}P_j\otimes {\rm e}^{\frac{z}{2}P_0}\sinh(\tfrac{z}{2}\eta J)\sin(\vartheta\eta J)\cos(\vartheta P_0)
\\
\phantom{\Delta_{\vartheta,z}(P_i)=}
{}-\eta  K_i\otimes {\rm e}^{\frac{z}{2}P_0}\sinh(\tfrac{z}{2}\eta J)\sin(\vartheta P_0)\cos(\vartheta\eta J)
\\
\phantom{\Delta_{\vartheta,z}(P_i)=}
{}-\eta \epsilon_{ij}K_j\otimes {\rm e}^{\frac{z}{2}P_0}\sinh(\tfrac{z}{2}\eta J)\left[\cos(\vartheta\eta J)\cos(\vartheta P_0)-1 \right],
\\
\Delta_{\vartheta,z}(K_i)
=\Delta_z(K_i)+{\rm e}^{-\frac{z}{2}P_0}\cosh(\tfrac{z}{2}\eta J)\left[\cos(\vartheta P_0)\cos(\vartheta\eta J)-1\right]\otimes K_i
\\
\phantom{\Delta_{\vartheta,z}(K_i)=}
{}+{\rm e}^{-\frac{z}{2}P_0}\cosh(\tfrac{z}{2}\eta J)\sin(\vartheta P_0)\cos(\vartheta\eta J)\otimes \epsilon_{ij}K_j
\\
\phantom{\Delta_{\vartheta,z}(K_i)=}
{}+{\rm e}^{-\frac{z}{2}P_0}\cosh(\tfrac{z}{2}\eta J)\frac{\sin(\vartheta\eta J)}{\eta}\cos(\vartheta P_0)\otimes P_i
\\
\phantom{\Delta_{\vartheta,z}(K_i)=}
{}+{\rm e}^{-\frac{z}{2}P_0}\cosh(\tfrac{z}{2}\eta J)\frac{\sin(\vartheta\eta J)}{\eta}\sin(\vartheta P_0)\otimes \epsilon_{ij}P_j
\\
\phantom{\Delta_{\vartheta,z}(K_i)=}
{}+K_i\otimes {\rm e}^{\tfrac{z}{2}P_0}\cosh(\tfrac{z}{2}\eta J) \left[\cos(\vartheta\eta J)\cos(\vartheta P_0)-1 \right]
\\
\phantom{\Delta_{\vartheta,z}(K_i)=}
{}-\epsilon_{ij}K_j\otimes {\rm e}^{\tfrac{z}{2}P_0}\cosh(\tfrac{z}{2}\eta J)\sin(\vartheta P_0)\cos(\vartheta\eta J)
\\
\phantom{\Delta_{\vartheta,z}(K_i)=}
{}-P_i\otimes {\rm e}^{\frac{z}{2}P_0}\cosh(\tfrac{z}{2}\eta J)\frac{\sin(\vartheta\eta J)}{\eta}\cos(\vartheta P_0)
\\
\phantom{\Delta_{\vartheta,z}(K_i)=}
{}+\epsilon_{ij}P_j\otimes {\rm e}^{\tfrac{z}{2}P_0}\cosh(\tfrac{z}{2}\eta J)\frac{\sin(\vartheta\eta J)}{\eta}\sin(\vartheta P_0)
\\
\phantom{\Delta_{\vartheta,z}(K_i)=}
{}-{\rm e}^{-\frac{z}{2}P_0}\sinh(\tfrac{z}{2}\eta J)\sin(\vartheta\eta J)\sin(\vartheta P_0)\otimes K_i
\\
\phantom{\Delta_{\vartheta,z}(K_i)=}
{}+{\rm e}^{-\frac{z}{2}P_0}\sinh(\tfrac{z}{2}\eta J)\sin(\vartheta\eta J)\cos(\vartheta P_0)\otimes \epsilon_{ij}K_j
\\
\phantom{\Delta_{\vartheta,z}(K_i)=}
{}+{\rm e}^{-\frac{z}{2}P_0}\frac{\sinh(\tfrac{z}{2}\eta J)}{\eta}\sin(\vartheta P_0)\cos(\vartheta\eta J)\otimes P_i
\\
\phantom{\Delta_{\vartheta,z}(K_i)=}
{}-{\rm e}^{-\frac{z}{2}P_0}\frac{\sinh(\tfrac{z}{2}\eta J)}{\eta}[\cos(\vartheta\eta J)\cos(\vartheta P_0)-1]\otimes\epsilon_{ij}P_j
\\
\phantom{\Delta_{\vartheta,z}(K_i)=}
{}+K_i\otimes {\rm e}^{\frac{z}{2}P_0}\sinh(\tfrac{z}{2}\eta J)\sin(\vartheta\eta J)\sin(\vartheta P_0)
\\
\phantom{\Delta_{\vartheta,z}(K_i)=}
{}+\epsilon_{ij}K_j\otimes {\rm e}^{\frac{z}{2}P_0}\sinh(\tfrac{z}{2}\eta J)\sin(\vartheta\eta J)\cos(\vartheta P_0)
\\
\phantom{\Delta_{\vartheta,z}(K_i)=}
{}+P_i\otimes {\rm e}^{\frac{z}{2}P_0}\frac{\sinh(\tfrac{z}{2}\eta J)}{\eta}\sin(\vartheta P_0)\cos(\vartheta\eta J)
\\
\phantom{\Delta_{\vartheta,z}(K_i)=}
{}+\epsilon_{ij}P_j\otimes {\rm e}^{\frac{z}{2}P_0}\frac{\sinh(\tfrac{z}{2}\eta J)}{\eta}\left[\cos(\vartheta\eta J)\cos(\vartheta P_0)-1\right],
\end{gather*}
for $i,j=1,2$.
Note that the limit $\vartheta\to 0$ is always well-def\/ined and gives the ``untwisted'' coproduct $\Delta_z$
in~\eqref{fa}.

\subsection{``Bicrossproduct-type'' basis}

Similarly to the previous subsection we now consider the $\kappa$-AdS$_\omega$ algebra expressed in the
``bi\-cross\-product-type'' basis introduced in~\cite{Rox}.
In this basis, the coproduct is given by
\begin{gather}
\Delta_z(P_0)=1\otimes P_0+P_0\otimes 1,
\qquad
\Delta_z(J)=1\otimes J+J\otimes 1,
\nonumber
\\
\Delta_z(P_i)={\rm e}^{-z P_0} \otimes P_i+P_i \otimes \cosh(z \eta J)-\eta \epsilon_{ij} K_j \otimes \sinh(z \eta J),
\nonumber
\\
\Delta_z(K_i)={\rm e}^{-z P_0} \otimes K_i+K_i \otimes \cosh(z \eta J)+\epsilon_{ij} P_j\otimes
\frac{\sinh(z \eta J)}{\eta},
\label{fe}
\end{gather}
and the deformed commutation rules read
\begin{gather*}
[J, P_i]= \epsilon_{ij} P_j,
\qquad
[J, K_i]= \epsilon_{ij} K_j,
\qquad
[J, P_0]= 0,
\qquad
[P_0, K_i]=-P_i,
\\
[P_0, P_i]=\omega K_i,
\qquad
 [P_1, P_2]=-\omega\frac{\sinh(2z\eta J)}{2 z \eta} ,
\qquad
[K_1, K_2]=-\frac{\sinh(2 z\eta J)}{2 z \eta},
\\
[P_i, K_j]
=\delta_{ij}\left\{\frac{{\rm e}^{-2 z P_0}-\cosh(2 z\eta J)}{2z}-\frac{\tan (z\eta)}{2\eta}\left({\mathbf P}^2+\omega {\mathbf K}^2\right)\right\}
+\frac{\tan (z\eta)}{\eta} (P_j P_i+\omega K_i K_j),
\end{gather*}
while the deformed Casimir invariants are given by
\begin{gather*}
{\cal C}=4 \cos (z\eta) \left\{ {\frac{\sinh^2(\frac z 2 P_0)}{z^2}}
\cosh^2\left(\frac z 2\eta J\right)+ {\frac{\sinh^2(\frac z 2 \eta J)}{z^2}}    \cosh^2\left(\frac z 2 P_0\right) \right\}
\\
\phantom{{\cal C}=}
{}-\frac{\sin (z\eta)}{z\eta}  {\rm e}^{z P_0}\left\{\cosh (z \eta J) \left(\mathbf{P}^2+\omega \mathbf{K}^2 \right)
-2\eta \sinh (z \eta J) (K_1 P_2-K_2 P_1) \right\},
\\
 {\cal W}=-\cos (z\eta)\frac{\sinh(z\eta J)}{z\eta}  \frac{\sinh(z P_0)}{z}
\\
\phantom{{\cal W}=}
{}+\frac{\sin (z\eta)}{z\eta}  {\rm e}^{z P_0}\left\{\cosh (z \eta J) (K_1 P_2-K_2 P_1)-\frac{\sinh (z \eta J)} {2
\eta} \left(\mathbf{P}^2+\omega \mathbf{K}^2 \right) \right\}.
\end{gather*}
By applying the twist~\eqref{twist} to~\eqref{fe} we obtain the two-parameter coproduct
\begin{gather*}
\Delta_{\vartheta,z}(P_0) = 1\otimes P_0+P_0\otimes 1,
\qquad
\Delta_{\vartheta,z}(J) = 1\otimes J+J\otimes 1,
\\
\Delta_{\vartheta,z}(P_i) = \Delta_z (P_i)+{\rm e}^{-zP_0}[\cos(\vartheta\eta J)\cos(\vartheta P_0)-1]\otimes P_i
+{\rm e}^{-zP_0}\sin(\vartheta P_0)\cos(\vartheta\eta J)\otimes \epsilon_{ij}P_j
\\
\qquad{}
-\eta {\rm e}^{-zP_0}\sin(\vartheta\eta J)\cos(\vartheta P_0)\otimes K_i-\eta {\rm e}^{-zP_0}\sin(\vartheta\eta
J) \sin(\vartheta P_0)\otimes \epsilon_{ij}K_j
\\
\qquad{}
+P_i\otimes \cosh(z\eta J) \left[\cos(\vartheta\eta J) \cos(\vartheta P_0)-1\right]-\epsilon_{ij}P_j\otimes
\cosh(z\eta J) \sin(\vartheta P_0) \cos(\vartheta\eta J)
\\
\qquad{}
+\eta K_i\otimes \cosh(z\eta J)\sin(\vartheta\eta J)\cos(\vartheta P_0)-\eta \epsilon_{ij}K_j\otimes \cosh(z\eta
J)\sin(\vartheta\eta J)\sin(\vartheta P_0)
\\
\qquad{}
+P_i\otimes \sinh(z\eta J)\sin(\vartheta\eta J)\sin(\vartheta P_0)+\epsilon_{ij}P_j\otimes \sinh(z\eta
J)\sin(\vartheta\eta J)\cos(\vartheta P_0)
\\
\qquad{}
-\eta  K_i\otimes \sinh(z\eta J)\sin(\vartheta P_0)\cos(\vartheta\eta J)-\eta \epsilon_{ij}K_j\otimes \sinh(z\eta
J)[\cos(\vartheta\eta J)\cos(\vartheta P_0)-1],
\\
\Delta_{\vartheta,z}(K_i)=\Delta_z (K_i)
+{\rm e}^{-zP_0}[\cos(\vartheta P_0)\cos(\vartheta\eta J)-1]\!\otimes
K_i\! +{\rm e}^{-zP_0}\sin(\vartheta P_0)\cos(\vartheta\eta J)\!\otimes  \epsilon_{ij}K_j
\\
\qquad{}
+{\rm e}^{-zP_0}\frac{\sin(\vartheta\eta J)}{\eta}\cos(\vartheta P_0)\otimes P_i+{\rm
e}^{-zP_0}\frac{\sin(\vartheta\eta J)}{\eta}\sin(\vartheta P_0)\otimes \epsilon_{ij}P_j
\\
\qquad{}
+K_i\otimes \cosh(z\eta J)[\cos(\vartheta\eta J)\cos(\vartheta P_0)-1]-\epsilon_{ij}K_j\otimes \cosh(z\eta
J)\sin(\vartheta P_0)\cos(\vartheta\eta J)
\\
\qquad{}
-P_i\otimes \cosh(z\eta J)\frac{\sin(\vartheta\eta J)}{\eta}\cos(\vartheta P_0)+\epsilon_{ij}P_j\otimes \cosh(z\eta
J)\frac{\sin(\vartheta\eta J)}{\eta}\sin(\vartheta P_0)
\\
\qquad{}
+K_i\otimes \sinh(z\eta J)\sin(\vartheta\eta J)\sin(\vartheta P_0)+\epsilon_{ij}K_j\otimes \sinh(z\eta J)\sin(\vartheta\eta J)\cos(\vartheta P_0)
\\
\qquad{}
+P_i\otimes \frac{\sinh(z\eta J)}{\eta}\sin(\vartheta P_0)\cos(\vartheta\eta J)+\epsilon_{ij}P_j\otimes
\frac{\sinh(z\eta J)}{\eta}[\cos(\vartheta\eta J)\cos(\vartheta P_0)-1].
\end{gather*}
Again, the untwisted coproduct~\eqref{fe} is straightforwardly recovered by setting~$\vartheta =0$.

As expected, there exists a~nonlinear mapping between the ``symmetrical'' and the ``bicrossproduct'' bases.
This is just the $(z, \vartheta)$-generalisation of the invertible nonlinear map introduced for $\kappa$-Poincar\'e
in~\cite{Majid:1994cy} and coincides with the one given in~\cite{Rox} for the quantum $\kappa$-AdS$_\omega$ algebra
without the $\vartheta$-twist.
If we denote by $\tilde Y_i$ the generators of the quantum twisted $\kappa$-AdS$_\omega$ algebra expressed in the above
``bicrossproduct-type'' basis and by~$Y_i$ the corresponding generators in the ``symmetrical'' basis of Section~\ref{Section5.1},
then the nonlinear map between both bases is given by
\begin{gather}
\tilde P_0=P_0,
\qquad
\tilde J= J,
\nonumber
\\
\tilde P_i={\rm e}^{-\frac z2 P_0}\left(\cosh\left(\frac z2 \eta J\right) P_i-\eta \sinh\left(\frac z2 \eta J\right)\epsilon_{ij}K_j \right) ,
\nonumber
\\
\tilde K_i={\rm e}^{-\frac z2 P_0}\left(\cosh \left(\frac z2 \eta J\right) K_i+ \frac{\sinh(\frac z2 \eta J)}{\eta}\epsilon_{ij} P_j \right).
\label{qa}
\end{gather}

\subsection[Quantum twisted $\kappa$-Poincar\'e algebra]{Quantum twisted $\boldsymbol{\kappa}$-Poincar\'e algebra}

The expressions above possess a~well-def\/ined limit $\omega=\eta^2\to 0$ which yields the corresponding structures for
the Poincar\'e algebra and can be compared with the results obtained in~\cite{Dasz1}.
In particular, the ``symmetrical'' basis gives rise to the twisted $\kappa$-Poincar\'e algebra with coproduct
\begin{gather*}
\Delta_{\vartheta,z}(P_0) = 1\otimes P_0+P_0\otimes 1,
\qquad
\Delta_{\vartheta,z}(J) = 1\otimes J+J\otimes 1,
\\
\Delta_{\vartheta,z}(P_i) = \Delta_z (P_i)+P_i\otimes {\rm e}^{\frac{z}{2} P_0}\left[\cos(\vartheta P_0)-1
\right]+{\rm e}^{-\frac{z}{2} P_0} \left[\cos(\vartheta P_0)-1\right]\otimes P_i
\\
\phantom{\Delta_{\vartheta,z}(P_i)=}
{}-\epsilon_{ij}P_j\otimes {\rm e}^{\frac{z}{2} P_0}\sin(\vartheta P_0)+{\rm e}^{-\frac{z}{2} P_0}\sin(\vartheta P_0)\otimes \epsilon_{ij}P_j,
\\
\Delta_{\vartheta,z}(K_i)=\Delta_z(K_i)+K_i\otimes{\rm e}^{\frac{z}{2}P_0}\left[\cos(\vartheta P_0)-1\right]
+{\rm e}^{-\frac{z}{2}P_0}\left[\cos(\vartheta P_0)-1\right]\otimes K_i
\\
\phantom{\Delta_{\vartheta,z}(K_i)=}
{}-\epsilon_{ij}K_j\otimes {\rm e}^{\frac{z}{2} P_0}\sin(\vartheta P_0)+{\rm e}^{-\frac{z}{2} P_0}\sin(\vartheta P_0)\otimes \epsilon_{ij}K_j
\\
\phantom{\Delta_{\vartheta,z}(K_i)=}
{}-\vartheta P_i\otimes {\rm e}^{\frac{z}{2} P_0}J\cos(\vartheta P_0)+\vartheta {\rm e}^{-\frac{z}{2}P_0}J\cos(\vartheta P_0)\otimes P_i
\\
\phantom{\Delta_{\vartheta,z}(K_i)=}
{}+\vartheta\epsilon_{ij}P_j\otimes{\rm e}^{\frac{z}{2} P_0}J\sin(\vartheta P_0)
+\vartheta{\rm e}^{-\frac{z}{2}P_0}J\sin(\vartheta P_0)\otimes\epsilon_{ij}P_j
\\
\phantom{\Delta_{\vartheta,z}(K_i)=}
{}+\frac{\displaystyle z}{2}P_i\otimes {\rm e}^{\frac{z}{2} P_0}J\sin(\vartheta P_0)
+\frac{\displaystyle z}{2}{\rm e}^{-\frac{z}{2} P_0}J\sin(\vartheta P_0)\otimes P_i
\\
\phantom{\Delta_{\vartheta,z}(K_i)=}
{}+\frac{\displaystyle z}{2}\epsilon_{ij}P_j\otimes {\rm e}^{\frac{z}{2} P_0}J[\cos(\vartheta P_0)-1]
-\frac{\displaystyle z}{2}{\rm e}^{-\frac{z}{2} P_0}J[\cos(\vartheta P_0)-1]\otimes\epsilon_{ij}P_j,
\end{gather*}
deformed commutation rules
\begin{gather*}
[J,P_i]=\epsilon _{ij}P_j,
\qquad
[J,K_i]=\epsilon _{ij}K_j,
\qquad
[J,P_0]=0,
\qquad
[P_0,P_i]=0,
\qquad
\left[P_0,K_i\right]=-P_i,
\\
\left[P_1,P_2\right]=0,
\qquad
\left[K_1,K_2\right]=-J\cosh(zP_0),
\qquad
\left[P_i,K_j\right]=-\delta_{ij}\frac{\displaystyle \sinh(zP_0)}{\displaystyle z},
\end{gather*}
and deformed Casimir operators
\begin{gather*}
\mathcal{C}=4\frac{\displaystyle \sinh^2(\tfrac{z}{2}P_0)}{\displaystyle z^2}-\mathbf{P}^2,
\qquad
\mathcal{W}=-J \frac{\displaystyle \sinh(zP_0)}{\displaystyle z}+K_1P_2-K_2P_1.
\end{gather*}
Performing the associated limit for the ``bicrossproduct-type'' basis, this gives rise to the quantum Poincar\'e algebra
with coproduct
\begin{gather*}
\Delta_{\vartheta,z}(P_0) =1\otimes P_0+P_0\otimes 1,
\qquad
\Delta_{\vartheta,z}(J) =1\otimes J+J\otimes 1,
\\
\Delta_{\vartheta,z}(P_i) = \Delta_z (P_i)+P_i\otimes [\cos(\vartheta P_0)-1]+{\rm e}^{-zP_0}[\cos(\vartheta P_0)-1]\otimes P_i
\\
\hphantom{\Delta_{\vartheta,z}(P_i) =}{}
-\epsilon_{ij}P_j\otimes \sin(\vartheta P_0)+{\rm e}^{-zP_0}\sin(\vartheta P_0)\otimes \epsilon_{ij}P_j,
\\
\Delta_{\vartheta,z}(K_i) = \Delta_z (K_i)+K_i\otimes [\cos(\vartheta P_0)-1]+{\rm e}^{-zP_0}[\cos(\vartheta P_0)-1]\otimes K_i
\\
\hphantom{\Delta_{\vartheta,z}(K_i) =}{}
-\epsilon_{ij}K_j\otimes \sin(\vartheta P_0)+{\rm e}^{-zP_0}\sin(\vartheta P_0)\otimes \epsilon_{ij}K_j-\vartheta P_i\otimes J\cos(\vartheta P_0)
\\
\hphantom{\Delta_{\vartheta,z}(K_i) =}{}
+\vartheta {\rm e}^{-zP_0}J\cos(\vartheta P_0)\otimes P_i
+\vartheta {\rm e}^{-zP_0}J\sin(\vartheta P_0)\otimes \epsilon_{ij}P_j+\vartheta \epsilon_{ij}P_j\otimes J\sin(\vartheta P_0)
\\
\hphantom{\Delta_{\vartheta,z}(K_i) =}{}
+zP_i\otimes J\sin(\vartheta P_0)+z\epsilon_{ij}P_j\otimes J[\cos(\vartheta P_0)-1],
\end{gather*}
commutation relations
\begin{gather*}
[J,P_i]=\epsilon_{ij}P_j,
\qquad
[J,K_i]=\epsilon_{ij}K_j,
\qquad
[J,P_0]=0,
\qquad
[P_0,P_i]=0,
\qquad
\left[P_0,K_i\right]=-P_i,
\\
\left[P_1,P_2\right]=0,
\qquad
\left[K_1,K_2\right]=-J,
\qquad
\left[P_i,K_j\right]=\delta_{ij}\left(\frac{\displaystyle{\rm e}^{-2zP_0}-1}{\displaystyle 2z}-\frac{\displaystyle z}{2}\mathbf{P}^2\right)+zP_jP_i,
\end{gather*}
and deformed Casimir operators
\begin{gather*}
\mathcal{C}=4\frac{\displaystyle \sinh^2(\tfrac{z}{2}P_0)}{\displaystyle z^2}-{\rm e}^{z P_0}\mathbf{P}^2,
\qquad
\displaystyle {{\cal W}=-J   \frac{\sinh(z P_0)}{z}+{\rm e}^{z P_0}\left(K_1 P_2-K_2 P_1-\frac z2  J
 \mathbf{P}^2 \right)}.
\end{gather*}
Note that the two bases are related via the Poincar\'e contraction $\eta\to 0$ of the map~\eqref{qa}, namely
\begin{gather*}
\tilde P_0=P_0,
\qquad
\tilde J= J,
\qquad
\tilde P_i={\rm e}^{-\frac z2 P_0} P_i,
\qquad
\tilde K_i={\rm e}^{-\frac z2 P_0}\left(K_i+\frac z2 \epsilon_{ij} J P_j \right) ,
\end{gather*}
and the correspondence with the results in~\cite{Dasz1} is obtained by setting $1/ {\kappa}=2 {\rm i}\vartheta $.

\section[Twisted $\kappa$-AdS$_\omega$ Poisson--Lie group]{Twisted $\boldsymbol{\kappa}$-AdS$\boldsymbol{_\omega}$ Poisson--Lie group}
\label{Section6}

It is well-known that, given a~classical~$r$-matrix $r= r^{ij}Y_i\otimes Y_j$ for a~Lie algebra $\mathfrak a$, the
Poisson--Lie (PL) brackets on the algebra of smooth functions on the associated PL group~$A$ are given by the Sklyanin
bracket~\cite{Drinfeldb}
\begin{gather}
\{f,g\}= r^{ij}\big(Y_i^Lf  Y_j^L g-Y_i^Rf  Y_j^R g\big),
\qquad
f,g\in {\rm Fun}(A),
\label{gb}
\end{gather}
where $Y_i^L$ and $Y_i^R$ are the left- and right-invariant vector f\/ields on~$A$.
In the case at hand, we have $\mathfrak a={\rm AdS}_\omega$, and $A={\rm SO}_\omega(2,2)$.
By inserting the vector f\/ields from Table~\ref{table1} and the classical~$r$-matrix~\eqref{ca} into~\eqref{gb}, we
obtain the PL brackets between the six {\em commutative} group coordinates $\{x_0,{\mathbf x},\theta,\boldsymbol{\xi}\}$ that
are dual to the basis elements $\{P_0, {\mathbf P}, J, {\mathbf K}\}$.
The PL subalgebra for the space-time local coordinates is given by
\begin{gather}
\{x_0,x_1\} =-z \frac{\tanh\eta x_1}{\eta \cosh^2\eta x_2}-\vartheta \cosh\eta x_1 \frac{\tanh\eta x_2}{\eta},
\nonumber
\\
\{x_0,x_2\} =-z \frac{\tanh\eta x_2}{\eta}+\vartheta  \frac{\sinh\eta x_1}{\eta},
\qquad
\{x_1,x_2\} =0,
\label{gc}
\end{gather}
and the ``crossed'' PL brackets read
\begin{gather}
 \{x_1,\xi_1\}
=\frac{z}{\cosh \eta x_2}\left(\frac{\cosh \eta x_2}{\cosh \eta x_1}-\frac{\cosh\xi_1}{\cosh \xi_2}+\tanh \eta x_1\sinh \eta x_2   A \right) ,
\nonumber
\\
 \{x_1,\xi_2\} =-z \cosh \xi_2  B,
\qquad
\{x_2,\xi_2\} = z\left(\frac{\cosh \eta x_1}{\cosh \eta x_2} \cosh \xi_1-\cosh \xi_2 \right) ,
\nonumber
\\
 \{x_2,\xi_1\} =-z A,
\qquad
\{\xi_1,\xi_2\} = z\eta \sinh \eta x_1 \left(C-\frac{\tanh \xi_2}{\cosh^2 \eta x_2} \right) ,
\nonumber
\\
 \{x_0,\theta\} =-\frac{z B}{\cosh \eta x_1}+\frac {\vartheta}2   \frac{\cosh \xi_1
\left(\cosh 2 \eta x_1-\cosh 2\xi_2 \right)}{\cosh \eta x_1 \cosh \eta x_2 \cosh \xi_2} ,
\nonumber
\\
 \{x_0,\xi_1\} = z\left(\frac{\sinh \xi_2}{\cosh \eta x_1} B-\frac{\sinh \xi_1\cosh \xi_2}{\cosh
\eta x_1\cosh \eta x_2} \right)-\vartheta  \frac{\cosh \eta x_1 \cosh \xi_1 \tanh \xi_2}{\cosh \eta x_2} ,
\nonumber
\\
 \{x_0,\xi_2\} =-z C+\vartheta \frac{\cosh \eta x_1\sinh \xi_1}{\cosh \eta x_2},
\qquad
\{\theta,x_1\} = z   \frac{\cosh \eta x_1}{\cosh \xi_2}  C+\vartheta  \frac{\sinh \xi_1\cosh \xi_2}{\cosh \eta x_2},
\nonumber
\\
 \{\theta,x_2\} =-z  \frac{\cosh \eta x_1\sinh \xi_1}{\cosh \eta x_2\cosh \xi_2}+\vartheta\sinh \xi_2 ,
\nonumber
\\
 \{\theta,\xi_1\} =-z\eta \left(\tanh \eta x_2+\tanh \eta x_1  B \right)
-\vartheta \frac{\eta \tanh \eta x_1\cosh \xi_1\cosh \xi_2}{\cosh \eta x_2} ,
\nonumber
\\
 \{\theta,\xi_2\} = \frac{z\eta \sinh \eta x_1}{\cosh^2 \eta x_2\cosh \xi_2}-\vartheta \eta \tanh
\eta x_2 \cosh \xi_2 ,
\label{ge}
\end{gather}
where the functions~$A$,~$B$ and~$C$ are given by
\begin{gather*}
 A=\frac {\sinh \eta x_1\sinh \eta x_2+\cosh \eta x_1\sinh \xi_1\tanh \xi_2} {\cosh \eta x_2} ,
\\
 B=\frac {\sinh \eta x_1\tanh \eta x_2 \cosh \xi_1+\sinh \xi_1\sinh \xi_2} {\cosh \eta x_2\cosh \xi_2} ,
\\
 C=\frac {\sinh \eta x_1\tanh \eta x_2 \sinh \xi_1+\cosh \xi_1\sinh \xi_2}{\cosh \eta x_1\cosh \eta x_2}.
\end{gather*}
Note {\sloppy that in lowest order, these PL brackets reduce to the f\/irst-order commutators~\eqref{cd} and~\eqref{ce}.
On the other hand, if the $\vartheta$-twist vanishes, we recover the PL brackets on ${\rm Fun}(\mbox{AdS$_\omega$})$
presented in~\cite{Rox}.

}

\subsection{Non-commutative (anti-)de Sitter and Minkowski spacetimes}

The f\/irst naive possibility for the quantisation of a~PL structure is given by the Weyl prescription and consists of
replacing the initial PL brackets between commutative group coordinates $y^i$ by Lie brackets between the corresponding
non-commutative coordinates $\hat y^i$.
This, indeed, works for linear PL structures, such as the $\kappa$-Poincar\'e
group~\cite{LukR,Majid:1994cy, Maslanka,kZakr}, in which the full set of PL brackets for the local spacetime coordinates
is linear in the deformation parameter $z=1/\kappa$.
However, for a~generic nonlinear PL bracket, this strategy will not work in general due to ordering ambiguities.

In the case of AdS$_\omega$, the PL brackets~\eqref{gc} and~\eqref{ge} should be quantised to give rise to the
non-commutative quantum group ${\rm Fun}_{z,\vartheta}(\mbox{AdS$_\omega$})$.
In particular, the non-commutative spacetime will arise as the quantisation of the Poisson algebra~\eqref{gc}.
In this case, although the brackets are nonlinear, we have $\{x_1,x_2\}=0$, and if we assume that the two corresponding
quantum coordinates commute $[\hat x_1, \hat x_2] =0$, the quantisation of the full PL bracket can be achieved via the
Weyl prescription.
As the only potential ordering ambiguities involve the coordinates $\hat x_1$ and $\hat x_2$, it follows that the
quantum twisted AdS$_\omega$ spacetime is given by
\begin{gather}
[\hat x_0, \hat x_1] =-z \frac{\tanh\eta \hat x_1}{\eta \cosh^2\eta \hat x_2}-\vartheta \cosh\eta \hat x_1
\frac{\tanh\eta \hat x_2}{\eta}
\nonumber
\\
\phantom{[\hat x_0, \hat x_1]}
=-z \left(\hat x_1-\frac13\omega\hat x_1^3-\omega\hat x_1\hat x_2^2\right)
-\vartheta\left(\hat x_2+\frac12\omega\hat x_1^2\hat x_2-\frac 13\omega \hat x_2^3 \right)+\mathcal{O}\big(\omega^2\big),
\nonumber
\\
[\hat x_0,\hat x_2] =-z \frac{\tanh\eta \hat x_2}{\eta}+\vartheta  \frac{\sinh\eta\hat x_1}{\eta}
\nonumber
\\
\phantom{[\hat x_0, \hat x_2]}
=-z\left(\hat x_2-\frac 13 \omega \hat x_2^3 \right)+\vartheta \left(\hat x_1+\frac 16 \omega \hat x_1^3 \right)
+\mathcal{O}\big(\omega^2\big),
\nonumber
\\
[\hat x_1, \hat x_2] =0.
\label{ha}
\end{gather}
Thus, we have shown how the f\/irst-order non-commutative space-time~\eqref{cd}, which is common to the {\em three}
quantum twisted $\kappa$-AdS$_\omega$ algebras, is generalised with an explicit dependence on the curvature or
cosmological constant $\omega=-\Lambda$.

The asymmetric form of~\eqref{ha} with respect to the $\hat x_1$ and $\hat x_2$ quantum coordinates could be expected
from the beginning, as we are dealing with {\em local} coordinates.
However, if we consider non-commutative ambient (Weierstrass) coordinates $(s_3, s_0,{{\mathbf s}})$, def\/ined in terms of $(x_0,
{{\mathbf x}})$ through~\eqref{bi}, we obtain that the PL twisted $\kappa$-AdS$_\omega$ spacetime can be written as a~quadratic
(and much more symmetric) Poisson algebra:
\begin{gather*}
\{s_0,s_i\}=-s_3 \left(z s_i+\vartheta \epsilon_{ij}s_j \right) ,
\qquad
\{s_1,s_2\}=0,
\\
\{s_3,s_0\}=z w   \mathbf{s}^2,
\qquad
\{s_3,s_i\}= w s_0 \left(z s_i+\vartheta \epsilon_{ij} s_j\right) ,
\end{gather*}
whose limit $\omega\to 0$ is given by $(s_3,s_0,{\mathbf s})\to (1, x_0, {{\mathbf x}})$.
In fact, the quantisation of this Poisson algebra in a~way consistent with the relations~\eqref{bi} will provide an
alternative description for the non-commutative spacetime~\eqref{ha}.

\section[Twisted $\kappa$-AdS quantum group]{Twisted $\boldsymbol{\kappa}$-AdS quantum group}
\label{Section7}

In this section, we relate the quantum deformation of ${\rm AdS}_\omega$ to the standard quantum deformations of
$\mathfrak{sl}(2,\mathbb R)$ via the Lie algebra isomorphism $\mathfrak{so}(2,2)\simeq\mathfrak{sl}(2,\mathbb{R})\oplus \mathfrak{sl}(2,\mathbb{R})$.
As in Section~\ref{Section3.4.1}, we therefore consider two copies of the Lie algebra $\mathfrak{sl}(2,\mathbb{R})$ with bases
$\{J_+^l,J_-^l, J_3^l\}$, group coordinates $(a_{+,l}, a_{-,l},\chi_l)$ and $(a_l,b_l,c_l,d_l)$ $(l=1,2)$.
To exhibit the relevant structures more clearly, we consider the {\em three-parametric}~$r$-matrix
\begin{gather}
\label{rmatrixabc}
r_{\alpha,\beta,\delta}= \alpha J_+^1\wedge J_-^1+\beta  J_+^2\wedge J_-^2+\tfrac{\delta}{2}  J_3^1\wedge J_3^2,
\end{gather}
which coincides with the classical~$r$-matrix $r_F'$ from~\eqref{rslsl} for $\alpha=\delta=-\frac \eta 2$ and
$\beta=\frac \eta 2$.
This allows one to determine the role of each term in the construction of the corresponding quantum AdS group.

\subsection[Quantum standard SL$(2,\mathbb{R})$ group]{Quantum standard SL$\boldsymbol{(2,\mathbb{R})}$ group}

To quantise the PL bracket def\/ined by the~$r$-matrix~\eqref{rmatrixabc}, it is worth recalling the well-known
construction of the standard quantum SL$(2,\mathbb{R})$ group (see, for instance,~\cite{Tak}).
For this, consider the Lie algebra $\mathfrak{sl}(2,\mathbb{R})$ with Lie bracket
\begin{gather*}
[J_{3},J_\pm]=\pm 2 J_\pm,
\qquad
[J_{+},J_{-}]=J_3,
\end{gather*}
and its fundamental representation given by
\begin{gather*}
J_3= \left(
\begin{matrix}
1&0
\\
0&-1
\end{matrix}
\right) ,
\qquad
J_+= \left(
\begin{matrix}
0&1
\\
0&0
\end{matrix}
\right) ,
\qquad
J_-= \left(
\begin{matrix}
0&0
\\
1 &0
\end{matrix}
\right).
\end{gather*}
This enables one to parametrise elements of the group ${\rm SL}(2,\mathbb{R})$ near the unit element according~to
\begin{gather}
T={\rm e}^{a_-J_-}{\rm e}^{a_+J_+}{\rm e}^{\chi J_3}= \left(
\begin{matrix}
{\rm e}^{\chi} & a_+ {\rm e}^{-\chi}
\\
a_- {\rm e}^{\chi} & (1+a_- a_+){\rm e}^{-\chi}
\end{matrix}
\right) \equiv \left(
\begin{matrix}
a & b
\\
c & d
\end{matrix}
\right) ,
\qquad
a~d-bc =1,
\label{kkb}
\end{gather}
where $(a_+,a_-,\chi)$ are local coordinates.
The corresponding left- and right-invariant vector f\/ields of ${\rm SL}(2,\mathbb{R})$ are given by
\begin{alignat}{3}
& Y_{J_+}^L={\rm e}^{2\chi} \partial_{a_+},
\qquad &&
Y_{J_-}^L=a_+^2 {\rm e}^{-2\chi} \partial_{a_+}+{\rm e}^{-2\chi} \partial_{a_-}+a_+ {\rm e}^{-2\chi} \partial_{\chi}, &
\nonumber
\\
& Y_{J_3}^L=\partial_{\chi},
\qquad &&
Y_{J_+}^R =(1+2 a_-a_+) \partial_{a_+}-a_-^2 \partial_{a_-}+a_- \partial_{\chi}, &
\nonumber
\\
& Y_{J_-}^R = \partial_{a_-},
\qquad &&
Y_{J_3}^R =-2 a_- \partial_{a_-}+2 a_+  \partial_{a_+}+\partial_{\chi}.&
\label{kkc}
\end{alignat}
The Sklyanin bracket~\eqref{gb} is induced by the standard (Drinfel'd--Jimbo)
classical~$r$-matrix~\cite{Drinfelda,Jimbo} given by
\begin{gather*}
r^{\rm DJ}= z J_+\wedge J_-,
\qquad
z={\rm e}^q,
\end{gather*}
which yields the following Sklyanin brackets in terms of the local coordinates $(a_+,a_-,\chi)$:
\begin{gather*}
\{\chi,a_+\}=-z  a_+,
\qquad
\{\chi,a_-\}=-z  a_-,
\qquad
\{a_+,a_-\}=-2z  a_-a_+.
\end{gather*}
Passing to the coordinates given by the matrix entries $(a,b,c,d)$ from~\eqref{kkb}, one f\/inds that this PL structure is
homogeneous quadratic
\begin{alignat}{4}
& \{b,a\}=z a b, \qquad && \{c,a\}=z a c, \qquad && \{c,b\}=0, &
\nonumber\\
& \{d,b\}=z b d, \qquad && \{d,c\}=z c d, \qquad && \{d,a\}=2z b c, &
\label{kkd}
\end{alignat}
and $C=a d-bc$ is a~Casimir function.

The quantisation of the PL algebra~\eqref{kkd} in terms of the non-commutative coordinates $(\hat a,\hat b,\hat c,\hat d)$ takes the form
\begin{alignat}{4}
& \hat b\hat a -q\hat a\hat b=0, \qquad && \hat c \hat a -q\hat a\hat c=0, \qquad && [\hat c,\hat b]=0, &
\nonumber
\\
& \hat d \hat b-q\hat b \hat d=0, \qquad && \hat d \hat c-q\hat c \hat d =0, \qquad && [\hat d,\hat a]=\big(q-q^{-1}\big)\hat b \hat c,&
\label{kkf}
\end{alignat}
and is compatible with the coproduct for the quantum ${\rm SL}(2,\mathbb{R})$ group that is induced by the group
multiplication $\Delta(\hat T)=\hat T  \dot \otimes\hat T$, namely
\begin{alignat}{3}
& \Delta(\hat a)=\hat a\otimes \hat a~+\hat b\otimes\hat c, \qquad && \Delta(\hat b)=\hat a\otimes \hat b+\hat b\otimes \hat d,&
\nonumber
\\
& \Delta(\hat c)=\hat c\otimes \hat a~+\hat d\otimes \hat c, \qquad && \Delta(\hat d)=\hat c\otimes \hat b+\hat d\otimes \hat d. &
\label{kkff}
\end{alignat}
Moreover, the deformed Casimir operator is just the ``quantum determinant'' of $\hat T$
\begin{gather*}
C_q= {\rm det}_q(T)=\hat a\hat d-q^{-1}\hat b\hat c.
\end{gather*}
Note that the commutation relations~\eqref{kkf} are consistent with the Poisson brackets~\eqref{kkd},
as we have $[\cdot,\cdot]=z\{\cdot,\cdot\}+\mathcal{O}[z^2]$.

\subsection[Quantisation of the AdS group in the $\mathfrak{sl}(2,\mathbb{R})\oplus \mathfrak{sl}(2,\mathbb{R})$ basis]
{Quantisation of the AdS group in the $\boldsymbol{\mathfrak{sl}(2,\mathbb{R})\oplus \mathfrak{sl}(2,\mathbb{R})}$ basis}

The PL brackets for the~$r$-matrix~\eqref{rmatrixabc} are readily obtained by inserting the vector f\/ields~\eqref{kkc}
into the Sklyanin bracket~\eqref{gb}.
A straightforward computation shows that, in terms of the SO$(2,2)$ coordinates $(a_l,b_l,c_l,d_l)$ $(l=1,2)$, this yields
\begin{alignat*}{3}
& \{a_1,b_1\}=-\alpha a_1 b_1,  \qquad&&  \{a_2,b_2\}=-\beta a_2 b_2, &
\\
& \{a_1,c_1\}=-\alpha a_1 c_1,  \qquad&&  \{a_2,c_2\}=-\beta a_2 c_2,&
\\
& \{a_1,d_1\}=-2\alpha b_1 c_1, \qquad&&  \{a_2,d_2\}=-2\beta b_2 c_2,&
\\
& \{b_1,c_1\}=0,          \qquad&& \{b_2,c_2\}=0,&
\\
& \{b_1,d_1\}=-\alpha b_1 d_1,  \qquad&&  \{b_2,d_2\}=-\beta b_2 d_2,&
\\
& \{c_1,d_1\}=-\alpha c_1 d_1,  \qquad && \{c_2,d_2\}=-\beta c_2 d_2,&
\end{alignat*}
which correspond to the terms with parameters $\alpha$ and $\beta$ in the~$r$-matrix~\eqref{rmatrixabc}.
The ``crossed terms'' are induced by the twist, which is labelled by the parameter $\delta$
\begin{alignat*}{3}
& \{a_1,a_2\}=0,         \qquad&& \{c_1,a_2\}=\delta c_1 a_2,&
\\
& \{a_1,b_2\}=-\delta a_1 b_2, \qquad && \{c_1,b_2\}=0,&
\\
& \{a_1,c_2\}=\delta a_1 c_2,  \qquad&&  \{c_1,c_2\}=0,&
\\
& \{a_1,d_2\}=0,        \qquad && \{c_1,d_2\}=-\delta d_2 c_1,&
\\
& \{b_1,a_2\}=-\delta b_1 a_2, \qquad && \{d_1,a_2\}=0,&
\\
& \{b_1,b_2\}=0,         \qquad&&  \{d_1,b_2\}=\delta d_1 b_2,&
\\
& \{b_1,c_2\}=0,         \qquad && \{d_1,c_2\}=-\delta c_2 d_1,&
\\
& \{b_1,d_2\}=\delta d_2 b_1,  \qquad && \{d_1,d_2\}=0.&
\end{alignat*}
It is immediate to check that $C_1=a_1 d_1-b_1 c_1$ and $C_2=a_2 d_2-b_2 c_2$ are Casimir functions for this
multi-parametric Sklyanin bracket.

The quantisation of this PL algebra is the quantum group ${\rm SO}_{q_\alpha,q_\beta,q_\delta}(2,2)$, where the
(non-intertwined) deformation parameters are $q_{\alpha}={\rm e}^{\alpha}$, $q_{\beta}={\rm e}^{\beta}$ and $q_{\delta}={\rm e}^{\delta}$.
Evidently, the quantum group coproduct is given by a~copy of~\eqref{kkff} for each of the two non-commutative sets of
coordinates $(\hat a_l,\hat b_l,\hat c_l,\hat d_l)$ ($l=1,2$).
The associated~$q$-commutation rules are the ones for two copies of the quantum $\mathrm{SL}(2,\mathbb R)$ group with
independent parameters
\begin{alignat*}{4}
&\hat b_1\hat a_1-q_{\alpha}\hat a_1\hat b_1=0,
\qquad&&
\hat c_1 \hat a_1-q_{\alpha}\hat a_1\hat c_1=0,
\qquad&&
[\hat c_1,\hat b_1]=0,&
\\
&\hat d_1 \hat b_1-q_{\alpha}\hat b_1 \hat d_1=0,
\qquad&&
\hat d_1 \hat c_1-q_{\alpha}\hat c_1 \hat d_1=0,
\qquad&&
[\hat d_1,\hat a_1]=\big(q_{\alpha}-q_{\alpha}^{-1}\big)\hat b_1 \hat c_1,&
\\
&\hat b_2\hat a_2-q_{\beta}\hat a_2\hat b_2=0,
\qquad&&
\hat c_2 \hat a_2-q_{\beta}\hat a_2\hat c_2=0,
\qquad&&
[\hat c_2,\hat b_2]=0,&
\\
&\hat d_2 \hat b_2-q_{\beta}\hat b_2 \hat d_2=0,
\qquad&&
\hat d_2 \hat c_2-q_{\beta}\hat c_2 \hat d_2=0,
\qquad&&
[\hat d_2,\hat a_2]=\big(q_{\beta}-q_{\beta}^{-1}\big)\hat b_2 \hat c_2.&
\end{alignat*}
Additionally, the quantum algebra exhibits ``crossed relations'' that are governed by the twist parameter:
\begin{gather*}
\left[\hat a_1,\hat a_2\right]=0, \qquad [\hat b_1,\hat b_2]=0, \qquad \left[\hat c_1,\hat c_2\right]=0, \qquad [\hat d_1,\hat d_2]=0,
\\
[\hat a_1,\hat d_2]=0,      \qquad [\hat a_2,\hat d_1]=0, \qquad [\hat b_1,\hat c_2]=0,      \qquad [\hat b_2,\hat c_1]=0,
\\
\hat a_2\hat b_1-q_{\delta}\hat b_1\hat a_2=0, \qquad \hat b_2\hat a_1-q_{\delta}\hat a_1\hat b_2=0,
\\
\hat a_1\hat c_2-q_{\delta}\hat c_2\hat a_1=0, \qquad \hat c_1\hat a_2-q_{\delta}\hat a_2\hat c_1=0,
\\
\hat b_1\hat d_2-q_{\delta}\hat d_2\hat b_1=0, \qquad \hat d_1\hat b_2-q_{\delta}\hat b_2\hat d_1=0,
\\
\hat c_2\hat d_1-q_{\delta}\hat d_1\hat c_2=0, \qquad \hat d_2\hat c_1-q_{\delta}\hat c_1\hat d_2=0.
\end{gather*}
To conclude the discussion, we consider the structure dual to the quantum group ${\rm SO}_{q_\alpha,q_\beta,q_\delta}(2,2)$
above, namely the Hopf algebra structure of the associated quantum algebra.
Its coproduct is given by a~formal series in the deformation parameters, and its f\/irst-order is the cocommutators
obtained via~\eqref{rcanon2} for the classical~$r$-matrix~\eqref{rmatrixabc}:
\begin{gather*}
\delta\big(J_3^1\big)=0,
\qquad
\delta\big(J_3^2\big)=0,
\\
\delta\big(J_+^1\big)=J_+^1\wedge \left(\alpha J_3^1-\delta J_3^2\right) ,
\qquad
\delta\big(J_+^2\big)=J_+^2\wedge \left(\beta J_3^2+\delta J_3^1\right) ,
\\
\delta\big(J_-^1\big)=J_-^1\wedge \left(\alpha J_3^1+\delta J_3^2\right) ,
\qquad
\delta\big(J_-^2\big)=J_-^2\wedge \left(\beta J_3^2-\delta J_3^1\right).
\end{gather*}
Thus, we are ef\/fectively dealing with two almost-disjoint copies of $\mathfrak{sl}(2,\mathbb{R})$ (not truly independent
due to the $\delta$-mixed terms in the cocommutators), and one readily obtains the full coproduct:
\begin{gather*}
\Delta\big(J_3^1\big)=J_3^1\otimes 1+1\otimes J_3^1,
\qquad
\Delta\big(J_3^2\big)=J_3^2\otimes 1+1\otimes J_3^2,
\\
\Delta\big(J_+^1\big)=J_+^1\otimes {\rm e}^{\frac{1}{2}(\alpha J_3^1-\delta J_3^2)}+{\rm e}^{-\frac{1}{2}(\alpha J_3^1-\delta J_3^2)}\otimes J_+^1,
\\
\Delta\big(J_-^1\big)=J_-^1\otimes {\rm e}^{\frac{1}{2}(\alpha J_3^1+\delta J_3^2)}+{\rm e}^{-\frac{1}{2}(\alpha J_3^1+\delta J_3^2)}\otimes J_-^1,
\\
\Delta\big(J_+^2\big)=J_+^2\otimes {\rm e}^{\frac{1}{2}(\beta J_3^2+\delta J_3^1)}+{\rm e}^{-\frac{1}{2}(\beta J_3^2+\delta J_3^1)}\otimes J_+^2,
\\
\Delta\big(J_-^2\big)=J_-^2\otimes {\rm e}^{\frac{1}{2}(\beta J_3^2-\delta J_3^1)}+{\rm e}^{-\frac{1}{2}(\beta J_3^2-\delta J_3^1)}\otimes J_-^2.
\end{gather*}
The deformed commutation relations for this quantum algebra can be calculated straightforwardly since the presence of
the twist $\frac \delta 2 J_3^1\wedge J_3^2$ in the classical~$r$-matrix~\eqref{rmatrixabc} does not af\/fect them.
Explicitly, they are given by
\begin{gather*}
\big[J_+^{l},J_-^{l}\big]=\frac{\sinh z_lJ_3^{l}}{z_l},
\qquad
\big[J_3^{l},J_{\pm}^{l}\big]=\pm 2J_{\pm}^l,
\qquad
l=1,2,
\end{gather*}
with $z_1=\alpha$ and $z_2=\beta$, while the basis elements of the dif\/ferent copies commute.

\section{Concluding remarks}

In this article, we have constructed the full quantum algebra as well as the associated non-commutative spacetimes for
the family of Lie algebras ${\rm AdS}_\omega$ with associated DD structures.
In these quantum algebras, the cosmological constant $\Lambda$ plays the role of a~deformation parameter in addition to
the energy scale given by $\kappa$, and a~further deformation parameter $\vartheta$ parametrises a~twist that is
motivated by the compatibility of this quantum deformation with (2+1)-gravity.

It would be interesting to investigate the impact of this twist in more detail by considering multi-particle models in
which the momenta are added via the coproduct of the quantum algebra.
While a~twist does not af\/fect the commutation relations of the quantum algebra, it manifests itself in the coproduct,
and, consequently, dif\/ferent values of the twist parameter $\vartheta$ lead to dif\/ferent momentum addition laws for
point particles.
It would be interesting to see if the precise value of this parameter that ensures the compatibility with (2+1)-gravity
is also motivated by physical considerations in the context of multi-particle models and whether it has a~geometrical
interpretation.

It would also be desirable to investigate in more depth the role of the cosmological constant or curvature in these
models and the spectra of the associated quantum operators.
At least in the AdS case, where the quantum algebra is obtained via a~twist from two commuting copies of the standard
quantum deformation of ${\rm SL}(2,\mathbb R)$, known results about the representation theory of this standard
deformation~\cite{CP, Drinfelda, Jimbo} should permit one to work out in detail the spectrum of the associated quantum
operators.

Finally, the introduction of another graded contraction parameter associated to the involution~$\Pi$ in~\eqref{inv}
would allow one to compute the non-relativistic limits of all the quantum twisted $\kappa$-AdS$_\omega$ algebras
presented in the paper.
This would lead to twisted $\kappa$-deformations of both the Newton--Hooke algebra and the Galilean one~\cite{Dasz1,Dasz2,DaszNH}.

\subsection*{Acknowledgements}

This work was partially supported by the Spanish MICINN under grant MTM2010-18556 and by the DFG Emmy-Noether fellowship
ME 3425/1-2.
P.N.~acknowledges a~postdoctoral fellowship from Junta de Castilla y Le\'on (Spain).

\pdfbookmark[1]{References}{ref}
\LastPageEnding

\end{document}